\documentclass[pre,eqsecnum,showpacs,amssymb]{revtex4}
\usepackage{amsmath,graphicx}
\usepackage{epsfig}

\begin{document}

\title{Forced-induced desorption of a polymer chain adsorbed on an attractive
surface - Theory and Computer Experiment}

\author{S. Bhattacharya$^1$, V. G. Rostiashvili$^1$, A. Milchev$^{1,2}$, and
T.A. Vilgis$^1$}
\affiliation{$^1$ Max Planck Institute for Polymer Research, 10 Ackermannweg,
55128 Mainz, Germany\\
$^2$ Institute for Physical Chemistry, Bulgarian Academy of Sciences, 1113
Sofia, Bulgaria
}

\begin{abstract}
We consider the properties of a self-avoiding polymer chain, adsorbed on a solid
attractive substrate which is attached with one end to a pulling force. The
conformational properties of such chain and its phase behavior are treated
within  a Grand Canonical Ensemble (GCE) approach. We derive theoretical
expressions for the mean size of loops, trains, and tails of an adsorbed chain
under pulling as well as values for the universal exponents which describe their
probability distribution functions. A central result of the theoretical analysis
is the derivation of an expression for the crossover exponent $\phi$,
characterizing polymer adsorption at criticality, $\phi = \alpha -1$, which
relates the precise value of $\phi$ to the exponent $\alpha$, describing polymer
loop statistics. We demonstrate that $1-\gamma_{11} < \alpha < 1 + \nu$,
depending on the possibility of a single loop to interact with neighboring
loops in the adsorbed polymer. The universal surface loop exponent $\gamma_{11}
\approx -0.39$ and the Flory exponent $\nu \approx 0.59$.

We present the adsorption-desorption phase diagram of a polymer chain under
pulling and demonstrate that the relevant phase transformation becomes first
order whereas in the absence of external force it is known to be a continuous
one. The nature of this transformation turns to be dichotomic, i.e., coexistence
of different phase states is not possible. These novel theoretical predictions
are verified by means of extensive Monte Carlo simulations.
\end{abstract}

\maketitle

\section{Introduction}

With the development of novel single macromolecule experiments, the manipulation
of individual polymer chains and biological macromolecules is becoming an
important method for understanding their mechanical properties and
characterizing the intermolecular interactions \cite{Strick,Celestini}. Much of
the related upsurge of interest into the statics and dynamics of single
macromolecules at surfaces has been spurred by the use of Atomic Force
Microscopy\cite{Hansma,Kikuchi,Rief,Kishino} (AFM) and optical/magnetic
tweezers\cite{Bustamante,Sviboda,Ashkin} which allow one to manipulate single
polymer chains.
Measurements of the force, needed to detach a chain from an adsorbing
surface, and most notably, of the force versus extension relationship which
exhibits sharp discontinuities have been interpreted as indication for the
presence of unadsorbed loops on the surface. In turn, this has initiated 
a number of theoretical studies \cite{Sevick,Livadaru,Serr} which have helped
to get better insight into the thermodynamic behavior and the mechanism of
polymer detachment from adhesive surface under pulling external force. A
comprehensive treatment of the problem for the case of a phantom polymer chain
can be found in the paper of Skvortsov et al. \cite{SKB}.
There is a close analogy between the forced detachment of an adsorbed polymer
chain like polyvinilamine and polyacrylic acid, adhering to a solid surface 
such as mica or a self-assembled monolayer, when the chain is pulled by the end
monomer, and the unzipping of homogeneous double-stranded DNA. In the context of
DNA denaturation and the simple single chain adsorption this analogy has been
discussed already in the middle 60s \cite{DiMarzio}. Recently, the DNA
denaturation and its unzipping  have been reconsidered by Kafri, Mukamel and
Peliti \cite{Kafri}. The consideration was based on the Poland and Sheraga' s
Grand Canonical Ensemble (GCE) approach \cite{Poland,Bir} as well as on Duplantier's
analysis of the number of configurations in polymer networks of arbitrary
topology \cite{Duplantier}. Duplantier's analysis makes it possible to calculate
the values of universal exponents which undergo renormalization due to excluded
volume effects. In particular, it has been shown by Kafri et al. \cite{Kafri}
that this renormalization procedure changes even the order of the  melting (or
denaturation) transition in DNA from second to first order.

In the present paper we use the approach of Kafri et al.\cite{Kafri} in order
to treat  the detachment of a  single chain from a sticky substrate when the
chain end is pulled by external force. It has been pointed out
earlier\cite{SKB} that the problem may be considered within the framework of two
different statistical ensembles, i.e., by keeping the pulling force {\em fixed}
while measuring the (fluctuating) position of the polymer chain end, or, by
measuring the (fluctuating) force necessary to keep the chain end at {\em fixed}
distance above the adsorbing plane. Our theoretical consideration
has been carried out in the fixed force ensemble whereas experimentalists
usually work in the fixed distance ensemble. We start in Section
\ref{sect_theor_noforce} with the consideration of the
conventional adsorption (i.e. force-free) problem where we derive a basic
expression for the {\em crossover exponent} describing polymer adsorption.
There we also consider theoretically some basic features of adsorbed polymer
chains as the variation of the average length of loops and tails in the chain
with changing strength of the adsorption potential. In Section
\ref{sect_theor_force} we extend our theoretical analysis to the case of
polymer adsorption in the presence of external force, and obtain results for
the main conformal properties of such chains as well as the relevant phase
diagram of the system. The properties of the simulation model are briefly
reviewed in Section \ref{sect_MC_model}, and then in Section \ref{sect_results}
we report on our most important results, gained in the course of the computer
experiment, and compare them to theoretical predictions. We end this work
in Section \ref{sect_summary} with a brief summary and discussion of the most
salient results of the present investigation.

\section{Single chain adsorption: loop-, train-, and tail statistics}
\label{sect_theor_noforce}

\begin{figure}[bht]
\includegraphics[scale=0.7]{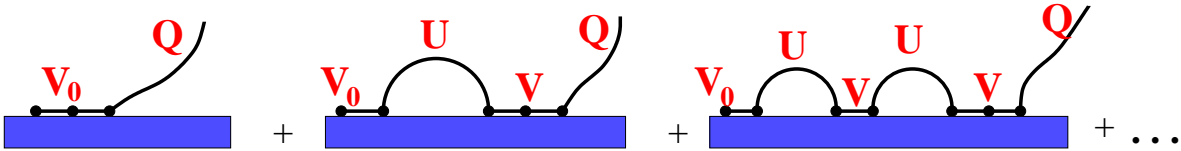}
\caption{Schematic representation of the series expansion given by Eq.
(\ref{GC_partition}) }
\label{Progression}
\end{figure}

A single chain, adsorbed on a solid plane, is built up from loops, trains and a
free tail. In order to derive expressions for the mean values of these basic
structural units, one may treat the problem within the Grand Canonical  Ensemble
(GCE)\cite{Bir}. In the GCE approach the lengths of these building blocks are not fixed
and are allowed to fluctuate. The GC-partition function is given as
\begin{eqnarray}
 \Xi (z) = \sum_{N=0}^{\infty} \: \Xi_{N} \: z^{N} = \frac{V_{0}(z) \: Q(z)}{1 -
V(z) U (z)}
\label{GC_partition}
\end{eqnarray}
where $\Xi_{N}$ is the canonical partition function of a chain of length $N$ and
$z$ is the fugacity. $U (z)$, $V (z)$ and $Q(z)$ denote the GC partition
functions of loops, trains and a tail respectively.  The building block
adjacent to the tethered chain end is allowed for by $V_{0} (z) = 1 + V(z)$.
The series given by Eq. (\ref{GC_partition}) is a geometric progression with
respect to $U(z) V(z)$. Figure \ref{Progression} gives a pictorial
representation of this series.

The GC-partition function of the loops is defined by
\begin{eqnarray}
 U (z)  = \sum_{m=1}^{\infty} \: \Omega_{m} \: z^{m} = \sum_{m=1}^{\infty} \:
\frac{(\mu_{3} z)^{m}}{m^{1-\gamma_{11}}}
\label{Loop}
\end{eqnarray}
where $\Omega_{m}$ is the number of surface $m$-loops (i.e., self-avoiding walks
of length $m$ which start and terminate on the surface) configurations. For an
isolated $m$-loop this number of configurations is given by $\Omega_{m} =
\mu_{3}^{m} m^{\gamma_{11}-1}$ where $\mu_{3}$ is the $3d$ connective constant
(in three dimensions, $d=3$, one has $\mu_{3} = 4.68$, and the exponent
$\gamma_{11} = -0.390$ )\cite{Vanderzande}. Below we will demonstrate that the
exponent $\gamma_{11}$ changes due to the excluded volume interactions between
different loops.

The train GC-partition function reads
 \begin{eqnarray}
V(z) =  \sum_{m=1}^{\infty} \: \Psi_{m} \: z^{m} = \sum_{m=1}^{\infty} \:
\frac{(\mu_{2} w  z)^m}{m^{1-\gamma_{d=2}}}
\label{Train}
 \end{eqnarray}
where the  number of train configurations of length $m$ (which are located in
the $d=2$ surface plane) is given by $\Psi_{m} = w^{m} \: \mu_{2}^{m} \;
m^{\gamma_{d=2} - 1}$. Here $\mu_2 = 2.6$ and $\gamma_{d=2} = 1.343$
\cite{Vanderzande}. In Eq.~(\ref{Train}) we have taken into account that each
adsorbed segment of the chain gains an additional statistical weight $w =
\exp(\varepsilon/k_BT)\equiv \exp(\epsilon)$, where $T$ is the temperature and
the Boltzmann constant $k_{B}$ is set to unity. In what follows the notation
$\epsilon$ stands for the dimensionless adsorption energy of a single monomer.
In fact, $\epsilon$ denotes the potential well depth of the short-ranged
surface potential, defined in the description of our simulation model in
Section~\ref{sect_MC_model}.

The GC-partition function for the chain tail is given by
\begin{eqnarray}
 Q (z) = 1 +  \sum_{m=1}^{\infty} \: \Lambda_{m} \: z^{m} = 1 +
\sum_{m=1}^{\infty} \: \frac{(\mu_{3} z)^m}{m^{1-\gamma_{1}}}
\label{Tail}
\end{eqnarray}
where the $m$-tail number of configuration equals $\Lambda_{m} = \mu_{3}^{m} \:
m^{\gamma_1 - 1}$, and in $d=3$ the exponent $\gamma_1 = 0.680$
\cite{Vanderzande}.

With the knowledge of the GC partition function, given by
Eq.(\ref{GC_partition}), it is possible to calculate the number of weighted
configurations of a polymer chain, containing $N$ segments (i.e., its canonical
partition function), $\Xi_{N}$. From the generating function method (see, e.g.,
Sec. $2.4$ in the book by Rudnick and Gaspari \cite{Rudnick}) it is well known
that at $N \rightarrow \infty$ the coefficient at $z^{N}$ is defined by a
singular point  (a pole or a branching point) of $\Xi(z)$ which lies closest to
the origin. In our case this is a simple pole, $z^{*}$, which is determined from
the condition
\begin{eqnarray}
V(z^*) \:  U(z^*) = 1
\label{Pole}
\end{eqnarray}
The principal contribution to this coefficient at $z^N$ is $(z^*)^{-(N+1)}$,
i.e., $\Xi_{N} \approx (z^*)^{-N}$, and so the corresponding free energy
\begin{eqnarray}
F = - T \ln \Xi_{N} = TN \ln z^*
\label{Free_Energy}
\end{eqnarray}
In Section \ref{sect_theor_noforce}, devoted to the adsorption of a pulled
polymer chain, we shall see that an important singularity arises also from the
tail generating function. The average fraction of adsorbed monomers, $n =
N_{s}/N$ (where $N_{s}$ is the number of adsorbed monomers) which we use as an
order parameter for the degree of adsorption, can be calculated then as follows
\begin{eqnarray}
 n \equiv \frac{N_{s}}{N} = \frac{1}{N} \; \frac{\partial \: \: \ln
\Xi_{N}}{\partial \ln w} = - \frac{\partial \ln z^*}{\partial \ln w}
\label{Fraction}
\end{eqnarray}

The generating functions, given by Eqs. (\ref{Loop}), (\ref{Train}), and
(\ref{Tail}), can be conveniently expressed in terms of the {\em polylog
function} \cite{Erdelyi}. In the Appendix we sketch the properties of the
polylog function and its behavior in the vicinity of the singular point.  In
terms of the polylog function (see Appendix) the basic Eq.(\ref{Pole}) is then
given by
\begin{eqnarray}
\Phi(\alpha, \mu_3 z^*) = \Phi^{-1}(\lambda, \mu_2 w z^*)
\label{Basic_Eq}
\end{eqnarray}
where the exponents $\alpha = 1 - \gamma_{11} \approx 1.39 > 1$ and $\lambda =
1-\gamma_{d=2} \approx - 0.343 < 1$. One should note that the exponent $\alpha =
1 - \gamma_{11}$ corresponds to a loop treated as an {\em isolated} one. This is
an important feature of the method which handles the main building blocks
(loops, trains and tails) as independent objects (see, e.g.,
Eq.(\ref{GC_partition})). Nevertheless, in Sec. \ref{Interaction}, following
Kafri et al. \cite{Kafri}, we shall show that by taking into account the
excluded
volume interaction between a loop and the rest of the chain  one ends up with a
renormalized value of the exponent $\alpha$ (it increases). This is important
because the value of $\alpha$ determines itself the value of the well known
surface (or, {\it crossover}) exponent $\phi$ in all the basic scaling laws
pertaining to polymer adsorption (see below).

\begin{figure}[bht]
\includegraphics[scale=0.4,angle=-90]{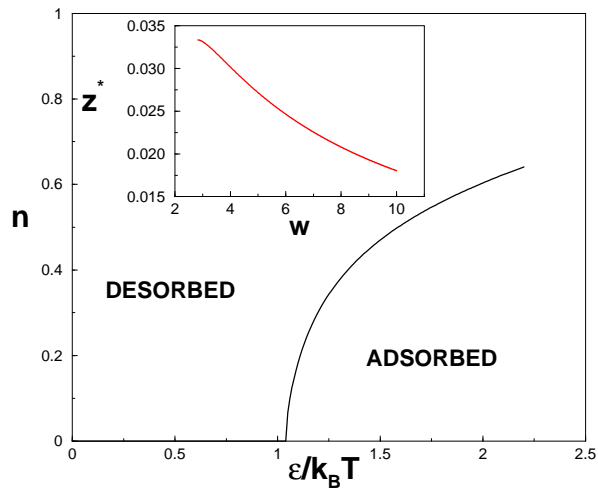}
\caption{The 'order parameter' (i.e., the fraction of adsorbed chain
segments), $n$, against the surface potential, $\epsilon$, in the absence of
detachment force, $f = 0$. The inset shows the variation of the fugacity $z^*$
with $w = \exp(\epsilon)$,  Eq. (\ref{Basic_Eq}).
}
\label{z_aster}
\end{figure}

Close to the critical point, $z_c = z^*$ which is defined by  $\mu_3 z_c =1$,
the l.h.s. of Eq.(\ref{Basic_Eq}) can be expanded (cf. Eq.(\ref{Cases}))  as
follows
\begin{eqnarray}
\zeta (\alpha) - a_{\alpha}(1 - \mu_3 z^*)^{\alpha - 1} - b_{\alpha} (1 - \mu_3
z^*) = \Phi^{-1}(1-\gamma_{d=2}, \mu_2 w z^*)
\label{Expansion}
\end{eqnarray}
with $\zeta(x)$ denoting the Riemann zeta-function.
At the critical adsorption point (CAP), $\epsilon_c$ and $w_c =
\exp(\epsilon_c)$, the solution of Eq. (\ref{Basic_Eq}) is $z^* = z_c = 1/\mu_3$
so that $w_c$ is given by the expression
\begin{eqnarray}
 \zeta (\alpha) = \Phi^{-1}(1-\gamma_{d=2}, \mu_2 w_c/\mu_3).
\label{Critical}
\end{eqnarray}
The expansion of Eq.(\ref{Expansion}) around the critical point, $z^{*} = z_c$
and $w = w_c$, could be effected by the substitution of $w = w_c + \delta$ and
$z^* = z_c - \Delta$ in Eq.  (\ref{Expansion}). Here  $\delta$ and $\Delta$ are
corresponding infinitesimal increments and we took into account that $z^*$
decreases with increasing $w$.  Substituting this in Eq. (\ref{Expansion}) gives
 \begin{eqnarray}
 \zeta (\alpha) - a_{\alpha}(\mu_3 \Delta)^{\alpha-1} \approx
\Phi^{-1}(1-\gamma_{d=2}, \mu_2 w_c z_c) - \Phi^{-2}(1-\gamma_{d=2}, \mu_2 w_c
z_c) \:\: \left[ \frac{d}{d x} \; \Phi (1-\gamma_{d=2},x)\right]_{x=\mu_2 w_c
z_c} \: \delta
\label{Expansion_1}
\end{eqnarray}
Taking into account the condition for the critical point, Eq. (\ref{Critical}),
as well as the identity Eq. (\ref{Property}), the solution for $z^*$ can be
recast in the form
\begin{eqnarray}
 z^*(w) \approx \frac{1}{\mu_3} \left[1 - \left(
\frac{A}{a_{\alpha}}\right)^{1/(\alpha-1)} \: (w - w_c)^{1/(\alpha-1)}\right]
\label{Solution}
\end{eqnarray}
 where the constants
\begin{eqnarray}
 A &=& \frac{\mu_2 \Phi(-\gamma_{d=2}, \mu_2 w_c/\mu_3)
}{\Phi^{2}(1-\gamma_{d=2}, \mu_2 w_c/\mu_3) } \nonumber\\
a_{\alpha}&=& \frac{\pi}{\Gamma(\alpha) |\sin (\pi \alpha)|},
\end{eqnarray}
and $w_c$ is defined by Eq. (\ref{Critical}). The full numerical solution for
the order parameter as well as for the pole $z^*(w)$ is displayed in Fig.
\ref{z_aster}.

Having the solution Eq.(\ref{Solution}) at hand, one can use the expression
Eq.(\ref{Fraction}) for the average fraction of adsorbed monomers. After some 
straightforward calculations one arrives at
\begin{eqnarray}
 n(\epsilon) \propto (\epsilon - \epsilon_c)^{\frac{1}{\alpha - 1} - 1}
\label{Staightforward}
\end{eqnarray}
where one has used  $w - w_c \approx \exp(\epsilon_c) \: (\epsilon -
\epsilon_c)$.
On the other hand,  it is well known \cite{Vanderzande} that the scaling
behavior in the vicinity of the critical adsorption energy is described by the
crossover exponent $\phi$. Namely, the corresponding scaling relationship is
given by
\begin{eqnarray}
 n(\epsilon) \propto \begin{cases}
                      N^{\phi - 1}, \quad &\mbox{at} \quad \epsilon =
\epsilon_c\\
(\epsilon - \epsilon_c)^{\frac{1}{\phi} - 1} \quad &\mbox{at} \quad \epsilon >
\epsilon_c
                  \end{cases}
\label{Scaling}
\end{eqnarray}
If the result, given by Eq.(\ref{Staightforward}), is compared to that of
Eq. (\ref{Scaling}), it becomes apparent that
\begin{eqnarray}
 \phi = \alpha -1
\label{Phi_Exponent}
\end{eqnarray}
This result, derived first by Birshtein\cite{Bir}, is of principal importance. 
Here it is derived in the context of self-avoiding chains. As stated
above, if the loops are treated as independent non-interacting objects, the
exponent $\alpha = 1 - \gamma_{11}$, so that
\begin{eqnarray}
  \phi = - \gamma_{11} \approx 0.39
\label{Phi_Exponent_0.39}
\end{eqnarray}
In Sec. \ref{Interaction} we shall demonstrate that by taking into account the
excluded volume interactions between a loop and the rest of the chain one finds
an increase of the values of $\alpha$, and $\phi$, respectively.

\subsection{Loops and tails distributions close to criticality}

Here we examine how the size distribution of polymer loops and tails looks like
close to the critical point of adsorption. The GC-partition function for loops,
given by Eq.(\ref{Loop}), yields immediately 
\begin{eqnarray}
P_{\rm loop} (l) \approx \left.  \frac{(\mu_3 z)^l}{l^{\alpha}}
\right|_{z=z^{*}}
= \frac{(\mu_3 z^{*})^l}{l^{1+\phi}}
\label{Loop_Distr}
\end{eqnarray}
where we have used the essential relation between the loop exponent $\alpha$ and
the crossover exponent $\phi$, Eq.(\ref{Phi_Exponent}). Close to the critical
point, $\mu_3 z^{*} \leq 1$ (see Eq.(\ref{Solution})) and the $l$-dependence is
mainly described by inverse power-law
\begin{eqnarray}
 P_{\rm loop} (l) \approx  \frac{1}{l^{1+\phi}}
\label{Inverse}
\end{eqnarray}
The power-law decay for the loop distribution close to the criticality has been
discussed in the early 80s by P.-G.de Gennes \cite{PGG}. Deeper in the region
of adsorption, however, the exponential part in Eq. (\ref{Loop_Distr})
dominates. Taking into account Eq.(\ref{Solution}), one obtains
\begin{eqnarray}
 P_{\rm loop} (l) \approx \frac{1}{l^{1+\phi}}\exp\left[ - c_1 (\epsilon -
\epsilon_c)^{1/\phi} \:
l\right],
\label{Exponent_law}
\end{eqnarray}
i.e., with increasing adsorption energy $\epsilon$ the size distribution becomes
narrower.

It is of interest to note that the size distribution of loops can be
reformulated in terms of the distribution $F(h)$ of projected lengths $h$
of the loops between two consecutive monomers residing on the adsorbing  surface
which was analyzed by Bouchaud and Daoud \cite{Bouchaud}. The relation between
$P_{\rm loop} (l)$ and $F(h)$ is straightforward, namely $F(h)dh = P_{\rm loop}
(l)|d l/d h| d h$, where due to the isotropy of loops $h \sim l^{\nu}$ too.
Taking these relations into account as well as Eq.(\ref{Inverse}) one obtains
\begin{eqnarray}
 F(h) \approx \frac{1}{h^{1+\phi/\nu}}
\label{Levy}
\end{eqnarray}
This corresponds exactly to the result of Bouchaud and Daoud \cite{Bouchaud}
where such broad distribution was associated with the so-called {\it
node-avoiding Levy flight}.

The distribution of tails (at the CAP, i.e., at $\mu_3 z^{*} \leq 1$) is even
broader, namely
\begin{eqnarray}
 P_{\rm tail} (l) \approx  \frac{(\mu_3 z^{*})^{l}}{l^{\beta}}
\label{Tail_Distribution}
\end{eqnarray}
where for an isolated tail $\beta = 1 - \gamma_1 \approx 0.32$.  We will show
below (see Eq.(\ref{Beta})) that if the interaction of a tail with the rest of
the chain is taken into account this leads to a larger value of $\beta = 0.51$.
One should be aware, however, that this result,
Eq.~(\ref{Tail_Distribution}), is only valid for $\epsilon \ge  \epsilon_c$
since a solution for Eq.~(\ref{Basic_Eq}) does not exist for subcritical values
of the adsorption potential. It is clear, however, that even in the subcritical
region,  $\epsilon <  \epsilon_c$, there are still monomers which occasionally
touch the substrate, creating thus single loops at the expense of the tail
length. This affects and modifies therefore the distribution  $P_{\rm tail}$ in
the vicinity of $\epsilon_c$. One can take into account this additional
contribution by considering a single loop - tail configuration. Pictorially
the latter can be inferred from Fig.~\ref{Contraction}b where instead of two
loops and a tail one should imagine a single loop adjacent to the tail. 
The partition function of such configuration is given by $Z_{l-t} =
\frac{\mu_3^{N-l}}{(N-l)^{1+\phi}}\;\frac{\mu_3^l}{l^\beta}$. On the other
side, the partition function of a tail conformation with no loops whatsoever
(i.e., of a tethered chain) is $Z_t = \mu_3^N\;N^{\gamma_1-1}$. Thus the
probability $P^<_{\rm tail}(l)$ to find a tail of length $l$ next to a single
loop of length $N-l$ can be estimated as
\begin{equation}
   P^<_{\rm tail}(l) = \frac{Z_{l-t}}{Z_t} \propto
\frac{N^{1-\gamma_1}}{l^\beta (N-l)^{1+\phi}}.
\label{Tail_distribution_below}
\end{equation}
Evidently, Eq.~(\ref{Tail_distribution_below}) predicts a singularity (that is,
a steep maximum) in the distribution of tails when $l \cong N$. One may expect
that in the vicinity of the critical point, $\epsilon \approx \epsilon_c$, the
observed distribution of tails will be given by an interpolation between the
expressions shown in  Eq.~(\ref{Tail_Distribution}) and
Eq.~(\ref{Tail_distribution_below}). Hence, the overall tail distribution can be
represented as
\begin{eqnarray}
P_{\rm tail}(l) =  \begin{cases}
                      \frac{1}{l^\beta}\exp\left [ -c_1 (\epsilon
-\epsilon_c)^{1/\phi}\;l\right ], \quad & \quad \epsilon > 
\epsilon_c\\
\\
\frac{A_1}{l^\beta} + \frac{A_2 N^{1-\gamma_1}}{l^\beta (N-l)^{1+\phi}}, \quad
& \quad \epsilon = \epsilon_c\\
\\
\frac{N^{1-\gamma_1}}{l^\beta (N-l)^{1+\phi}}. \quad & \quad  \epsilon <
\epsilon_c
                  \end{cases}
\label{Tail_distribution__all}
\end{eqnarray}
Evidently, close to the CAP this distribution is expected to attain a
$U$-shaped form with maxima at $l\approx 1$ and $l\cong N$. This shape of
$P_{\rm tail}(l)$ has been predicted earlier for a Gaussian chain by Gorbunov
et al.\cite{Gorbunov}. In close analogy with
Eq.~(\ref{Tail_distribution__all}), the distribution of loops reads
\begin{eqnarray}
P_{\rm loop}(l) =  \begin{cases}
                      \frac{1}{l^{1+\phi}}\exp\left [ -c_1 (\epsilon
-\epsilon_c)^{1/\phi}\;l\right ], \quad & \quad \epsilon > 
\epsilon_c\\
\\
\frac{B_1}{l^{1+\phi}} + \frac{B_2 N^{1-\gamma_1}}{l^{1+\phi}(N-l)^\beta},
\quad
& \quad \epsilon = \epsilon_c\\
\\
\frac{N^{1-\gamma_1}}{l^{1+\phi}(N-l)^\beta}. \quad & \quad  \epsilon <
\epsilon_c
                  \end{cases}
\label{Loop_distribution_all}
\end{eqnarray}
In Eqs.~(\ref{Tail_distribution__all})-(\ref{Loop_distribution_all}) $A_1, A_2,
B_1, B_2$ are some constants. As we shall see in Section \ref{sect_results},
the simulation results for $P_{\rm tail}(l), P_{\rm loop}(l)$ are in good
agreement with the predictions,
Eqs.~(\ref{Tail_distribution__all})-(\ref{Loop_distribution_all}).

\subsubsection{Divergence of the average loop and tail lengths at
 criticality}\label{Diverges}

The average loop length is defined by the loop GC-partition function,
Eq.(\ref{Loop}), as
\begin{eqnarray}
 L = \left. z \:\frac{\partial \ln U (z)}{\partial z}\right|_{z=z^*} =
\frac{\Phi (\alpha-1, \mu_{3} z^*)}{\Phi (\alpha, \mu_{3} z^*)}
\label{L}
\end{eqnarray}
where we have used Eq.(\ref{Property}).
Taking into account the polylog function behavior given by Eq.(\ref{Cases}) with
the requirement that $1 < \alpha < 2$ as well as the solution for $z^*$,
Eq.(\ref{Solution}), one gets
\begin{eqnarray}
 L \approx \frac{\Gamma(2-\alpha)}{\zeta(\alpha)} \: \left(
\frac{a_{\alpha}}{A}\right)^{\frac{2-\alpha}{\alpha-1}} \: \frac{1}{(w -
w_c)^{\frac{2-\alpha}{\alpha-1}}}\propto \frac{1}{(\epsilon -
\epsilon_c)^{\frac{1}{\phi} - 1}}
\label{L_diverge}
\end{eqnarray}
where the result Eq.(\ref{Phi_Exponent}) has been used. This result is
compatible with the scaling prediction based on Eq. (\ref{Scaling}). Indeed,
close to criticality, $L \approx N/N_s$. From Eq.(\ref{Scaling}) one obtains
then the same result, $L \propto (\epsilon - \epsilon_c)^{1 - 1/\phi}$. The free
energy goes as $F = T N \ln z^* \propto - N(\epsilon - \epsilon_c)^{1/\phi}$
where one has used Eq.(\ref{Solution}). On the other hand, the free energy is
proportional to the number of adsorption blobs, i.e., $F \propto N/g$, where $g$
is the length (number of segments) of the blob. The adsorption blobs are defined
to contain as many monomers $g$ as necessary to be on the verge of  adsorption
and therefore carry an adsorption energy of the order of $k_B T$ each. In result
the blob length scales as $g \propto (\epsilon - \epsilon_c)^{-1/\phi}$.  The
size of the adsorbed chain perpendicular to the surface, $R_{\perp}$, is nothing
 but the blob size, that is, $R_{\perp} \approx g^{\nu}$. Thus one obtains
\begin{eqnarray}
R_{\perp} \propto \frac{1}{(\epsilon - \epsilon_c)^{\nu/\phi}}.
\label{Blob_Size}
\end{eqnarray}

Consider now the average tail length $S$. In terms of the GC-partition
function for tails, Eq. (\ref{Tail}), it reads
\begin{eqnarray}
 S = \left. z \: \frac{\partial \ln Q(z)}{\partial z}\right|_{z=z^*} =
\frac{\Phi (\beta - 1, \mu_{3} z^*)}{1 + \Phi (\beta, \mu_3 z^*)}
\label{T}
\end{eqnarray}
with the exponent $\beta = 1 - \gamma_{1} = 0.32 < 1$. This value of the
exponent $\beta$ does not allow for the interaction of the tail with other
building blocks of the adsorbed chain and will be corrected in
Sec.\ref{Interaction}. Using the results, Eqs. (\ref{Cases}) and
(\ref{Solution}), the expression for the average tail length Eq.(\ref{T}) can be
recast in the form
\begin{eqnarray}
 S \approx (1 - \beta) \left( \frac{a_{\alpha}}{A}\right)^{\frac{1}{\alpha-1}}
\:
\frac{1}{(w - w_c)^{\frac{1}{\alpha - 1}}}\propto \frac{1}{(\epsilon -
\epsilon_c)^{\frac{1}{\phi}}}
\label{Tail_Length}
\end{eqnarray}
Notably, the exponent $\beta$ drops out of this expression. The corresponding
tail size $R_{\rm S} \sim S^{\nu}$ scales as
\begin{eqnarray}
 R_{\rm S} \propto \frac{1}{(\epsilon - \epsilon_c)^{\nu/\phi}}
\label{Tail_Size}
\end{eqnarray}
Note that the tail size, Eq.(\ref{Tail_Size}), scales exactly like the blob (and
not the loop!) size, Eq. (\ref{Blob_Size}).

\subsection{Role of {\em interacting} loops and tails}\label{Interaction}

As mentioned above, the exponent $\alpha$, which governs the numbers of loops in
the configuration of adsorbed polymer, determines also the crossover exponent
$\phi$ so that it is of prime importance to know the exact value of $\alpha$. If
the surface loops are treated as isolated objects (i.e., loop-loop or loop-tail
interactions are ignored), the exponent $\alpha = 1 - \gamma_{11} = 1.39$.
Recently Kafri {\it et al.} \cite{Kafri} have shown in the context of DNA
melting that the interaction of a loop with the rest of the chain increases the
loop exponent $\alpha$. In their work the authors of ref. \cite{Kafri}
essentially used some results of the renormalization theory of arbitrary polymer
graphs, developed earlier by Duplantier \cite{Duplantier}. This approach makes
it possible to treat also polymer chains which are grafted onto a solid surface.
Here we give a short sketch of Duplantier's results for a polymer graph located
close to the surface and then demonstrate how the loop-loop and loop-tail
interactions lead to the enhancement of the effective surface loop exponent.

\begin{figure}[bht]
\includegraphics[scale=0.7]{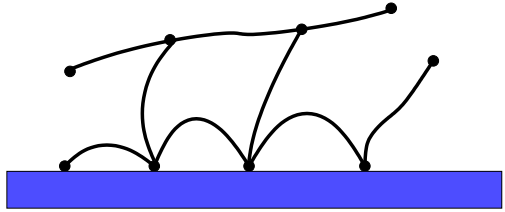}
\caption{A polymer graph located close to the surface has the following
topological characteristics: ${\cal L} = 1$, ${\cal L}_{s} = 9$, ${\cal V}_{s} =
4$, $n_1 = 3$, $n_3 = 2$,  $n_1^s = 1$, $n_2^s = 1$, $n_3^s = 2$. One surface
vertex is fixed whereas the other vertices may move freely.}
\label{Example}
\end{figure}

For an arbitrary self-avoiding polymer graph ${\cal G}$, which is grafted on the
surface, it has been shown \cite{Duplantier} that the total number of
configurations is given by the standard asymptotic expression:
\begin{eqnarray}
 Z({\cal G})= \mu_{3}^{N} \: N^{\gamma^{s} - 1}
\label{Number}
\end{eqnarray}
where $N = \sum_{j=1}^{\cal N} M_j$ is the total length of the graph made of
${\cal N}$ chains (or edges) of length $M_j$. The surface exponent $\gamma^{s}$
is given by the
following general relationship
\begin{eqnarray}
 \gamma^{s} =  1- \nu (d {\cal L} + {\cal L}_{s} + {\cal V}_s - 1) +
\sum\limits_{k\geq 1} \: (n_L \sigma_k + n_{k}^{s} \sigma_{L}^{s})
\label{Gamma_Exponent} 
\end{eqnarray}
where $\nu$ is the Flory exponent and $d$ stands for the space dimensionality.
In Eq. (\ref{Gamma_Exponent}) ${\cal L}$ is the total number of independent
constitutive polymer loops in the graph ${\cal G}$ (i.e., the surface loops are
not included in ${\cal L}$). ${\cal L}_{s}$  is the total number of extremities
of polymer lines upon contact to the surface. $n_k$ and $n_{k}^{s}$ are the
numbers of bulk and surface vertices of order $k$ respectively, thus ${\cal
L}_{s} = \sum_{k \geq 1} k \: n_{k}^{s}$. ${\cal V}_{s}$ gives the number of  
surface vertices, i.e., ${\cal V}_{s} = \sum_{k \geq 1} n_{k}^{s}$.
Finally, $\sigma_{k}$ and $\sigma_{k}^{s}$ are critical bulk and surface
exponents which correspond to the $k$-arm  vertices.  In $d < 4$ these exponents
can be calculated analytically via the $\varepsilon$-expansion but some of them
could be also expressed in terms of the conventional exponents $\nu$, $\gamma$,
$\gamma_1$ and $\gamma_{11}$ \cite{Duplantier}. Figure \ref{Example} gives an
example of a polymer graph with the specification of its topological elements.

The number of configurations given by Eq.(\ref{Number}) holds when the lengths
of all components $M_a$ are large and comparable to the total length $N$. As
long as at least one of them becomes  small, i.e. $M_a \ll N$,  then one gets
\begin{eqnarray} 
Z({\cal G})= \mu_{3}^{N} \: N^{\gamma^{s} - 1} \: G\left( \frac{M_1}{N},
\frac{M_2}{N}, \dots, \frac{M_{\cal N}}{N}\right)
\label{Function_G}
\end{eqnarray}
where the scaling function $G(x_1, x_2, \dots, x_{\cal N})$ has a singularity,
provided any of the arguments $x_a$ goes to zero. In fact, in this limit the
polymer graph changes its topology and, therefore, the  surface exponent
$\gamma^{s}$ changes too. In the next subsection we show how these results could
be used  to calculate the effective exponent $\alpha$ which takes into account
the interaction of a surface loop with the rest of the chain.

\subsubsection{Surface loop embedded in an adsorbed chain}

\begin{figure}[bht]
\includegraphics[scale=0.5]{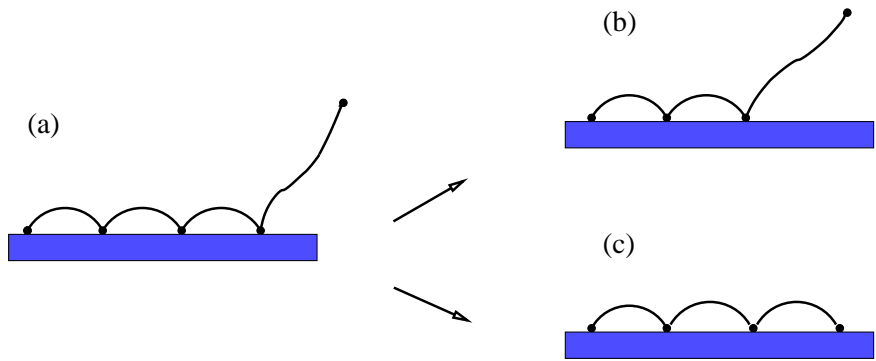}
\caption{An array of surface loops close to criticality. One of the surface
loops of length $M$ in the limit $M/N \ll 1$ is contracted, changing  the
topology of the polymer graph from (a) to (b). This contraction procedure makes
it possible to derive the scaling function $G(x)$. By similar contraction of a
tail the graph  goes over from (a) to (c).}
\label{Contraction}
\end{figure}

Consider the configurations of a chain (tethered with one end on the surface) in
the vicinity of the adsorption critical point (see Fig. \ref{Contraction}). Let
$M$ be the length of a surface loop while $K$ measures the length of the rest of
the chain, i.e., $M + K = N$. The number of configurations of the polymer graph,
depicted in Fig.\ref{Contraction}a, is
\begin{eqnarray}
 Z = \mu_{3}^{M+K} \: (M + K)^{\gamma_{a}^{s} - 1} \: G\left(
\frac{M}{M+K}\right)
\end{eqnarray}
where $\gamma_{a}^{s}$ is the exponent which could be calculated using Eq.(
\ref{Gamma_Exponent}) (see below) and the scaling function  $G(x)\approx 1$  for
large $M$ and $K$. In the case when $M/K \rightarrow 0$ one has a crossover to
the polymer graph shown in Fig \ref{Contraction}b where the number of
configurations $Z \sim \mu_{3}^{K} \: ( K)^{\gamma_{a}^{s} - 1} \:
(1/K)^{\gamma_a^s - \gamma_b^s}$ (with $\gamma_b^s$ being the surface exponent
of the corresponding graph). These arguments fix the form of the scaling
function which can be written as
\begin{eqnarray}
 G(x) \approx \begin{cases}
              x^{\gamma_a^s - \gamma_b^s}, \quad &\mbox{at}  \quad x \ll 1\\
1 , \quad &\mbox{at}  \quad x \approx 1
             \end{cases}
\label{G}
\end{eqnarray}

In the case of a  small surface loop, embedded in an adsorbed polymer, with  $M
\gg 1$ and $K\gg 1$ (but with $M/K \ll 1$) one obtains for the total number of
configurations 
 \begin{eqnarray}
  Z \sim \mu_{3}^{M+K} \: K^{\gamma_{a}^{s} - 1}
\left(\frac{M}{K}\right)^{\gamma_a^s - \gamma_b^s} \sim \mu_{3}^{M} \:
M^{\gamma_a^s - \gamma_b^s} \:\: \mu_{3}^{K} \: K^{\gamma_b^s - 1}.
 \label{Factorisation}
\end{eqnarray}
The last result indicates that the total partition function may be factorized to
$Z \sim Z_{\rm loop} \: Z_{\rm rest} $  where $Z_{\rm loop}$ and $Z_{\rm rest}$
are the partition functions of the small loop and the rest of the chain,
respectively. Thus, using the notations of Eq.(\ref{Loop}), one obtains
$\Omega_n = \mu_{3}^{n}/n^{\gamma_b^s - \gamma_a^s}$, i.e., the effective
exponent $\alpha$ becomes
 \begin{eqnarray}
 \alpha = \gamma_b^s - \gamma_a^s.
\label{Effective_alpha}
 \end{eqnarray}

Now we are in a position to determine the exponents $\gamma_a^s$ and
$\gamma_b^s$.  Let us assume that the polymer graph in Fig. \ref{Contraction}a
is made of ${\cal N}$ subchains (${\cal N} - 1$ being loops and $1$ - a tail).
The topological characteristics can be specified as follows: ${\cal L}=0$,
${\cal L}_s = 2{\cal N} - 1$, ${\cal V}_s = {\cal N}$, $n_1 = 1$, $n_1^s = 1$,
$n_2^s = {\cal N} - 1$. Earlier it has been shown \cite{Duplantier} that the
critical exponent $\sigma_2^s = 2\nu - 1$.  With these values
Eq.(\ref{Gamma_Exponent}) yields
\begin{eqnarray}
 \gamma_a^s = 2  - {\cal N}(\nu + 1) + \sigma_1 + \sigma_1^s
\label{Gamma_A}
\end{eqnarray}
The corresponding expression for $\gamma_b^s $ can be obtained from Eq.
(\ref{Gamma_A}) by the substitution ${\cal N} \rightarrow {\cal N} - 1$. This
yields
\begin{eqnarray}
 \gamma_b^s = 3  + \nu - {\cal N}(\nu + 1) + \sigma_1 + \sigma_1^s
\label{Gamma_B}
\end{eqnarray}

The final expression for the exponent $\alpha$,  given by Eq.
(\ref{Effective_alpha}), then reads
\begin{eqnarray}
 \alpha = \gamma_b^s - \gamma_a^s = \nu + 1.
\label{Final_Alpha}
\end{eqnarray}
With this theoretical prediction the value of the crossover exponent, given
by Eq. (\ref{Phi_Exponent}),  is determined as:
\begin{eqnarray}
 \phi = \alpha - 1 = \nu = 0.588
\label{Exponent_Nu}
\end{eqnarray}
where we have taken the best numerical estimate for the Flory exponent $\nu$ at
$d = 3$ \cite{Vanderzande}.
A comparison of Eq.(\ref{Exponent_Nu}) with Eq. (\ref{Phi_Exponent_0.39}) leads
to the important conclusion that, depending on the range of the excluded volume
interaction, the value of $\phi$ may vary significantly. Indeed, if the
interactions affect beads from the same surface loop only then $\phi$ is given
by Eq.(\ref{Phi_Exponent_0.39}),  otherwise (i.e., when the beads from all loops
interact) the value of $\phi$ will be enhanced markedly (see Eq.
(\ref{Exponent_Nu})).

One should emphasize, however, that Eq.~(\ref{Final_Alpha}) does not give an
{\em exact} value for the exponent $\alpha$, but rather an upper limit only.
Indeed, the total number of configurations, given by Eq.~(\ref{Factorisation}), 
is estimated by a {\em factorized} expression for the partition function which
takes into account the contribution of a loop and the rest of the chain. As a
 matter of fact this is a Mean Field approach which overestimates interactions
at the expense of correlations, reducing thus the total number of
configurations of a loop. The latter is reflected by an increase of $\alpha$.
The precise value of $\alpha$ therefore satisfies the inequality $1-\gamma_{11}
< \alpha < 1 + \nu$. In the special case of a Gauissian chain both the lower
and upper limits for $\alpha$ merge while for a phantom chain one
has $\gamma_{11} = -0.5$  (cf. Section 6.2 in \cite{Duplantier}) and $\nu =
0.5$. Thus, for Gaussian chains one obtains the well known value $\phi = 0.5$.

Following the same way of reasoning, one may expect that the exponent for the
tail, $\beta$ (see Eq. (\ref{T})), is also renormalized due to interaction with
the rest of the adsorbed chain. Tail contraction when going from (a) to (c) in
Fig. \ref{Contraction} enables one to obtain for the renormalized
$\beta$-exponent the following relationship
\begin{eqnarray}
 \beta = \gamma_{c}^s - \gamma_{a}^s
\label{Tail_Exponent}
\end{eqnarray}
where $\gamma_{c}^s$ is the surface exponent of the polymer graph given in Fig.
\ref{Contraction}c.  Again, if the polymer graph given in Fig.
\ref{Contraction}a is made of ${\cal N}$ chains then the exponent $\gamma_{c}^s$
for the graph Fig. \ref{Contraction}c becomes
\begin{eqnarray}
 \gamma_{c}^s = 3 - \nu - {\cal N}(\nu + 1) + 2 \sigma_{1}^s
\label{Gamma_C}
\end{eqnarray}
Taking into account Eq. (\ref{Gamma_A}), one obtains $\beta = 1 - \nu +
\sigma_1^s - \sigma_1$, whereby the critical exponents (see \cite{Duplantier})
are given by
\begin{eqnarray}
 \sigma_1 &=& \frac{\gamma - 1}{2}\nonumber\\
\sigma_1^s &=& \nu + \gamma_1 - \frac{\gamma + 1}{2}
\end{eqnarray}
The calculation gives finally 
\begin{eqnarray}
 \beta = \gamma_1 - \gamma + 1
\label{Beta}
\end{eqnarray}
with $\gamma_1 \approx 0.68$ and $\gamma \approx 1.17$ so that $\beta \approx
0.51$. As expected, the value of the $\beta$-exponent {\em increases} as
compared to the ``isolated tail'' case, $\beta = 1 - \gamma_{1} \approx 0.32$.

\subsubsection{Comparison with other results}

The result, given by Eq. (\ref{Exponent_Nu}), deserves a more detailed
discussion. One should point out that, generally, the value of $\phi$ for the
good solvent case in three dimension has been so far fairly controversial. For
example, Monte-Carlo (MC) data (albeit for relatively short chains $N \leq 100$)
on a diamond lattice yield $\phi = 0.588 \pm 0.03$ \cite{Eisenriegler} which is
in complete agreement with Eq.(\ref{Exponent_Nu}). A recent MC-investigation
\cite{Descas} has suggested that the uncertainty in the value of $\phi$ might be
related to the limited accuracy in the determination of the critical adsorption
energy $\epsilon_c$. Namely, for the bond fluctuation model (BFM), which has
been used by Descas, Sommer and Blumen \cite{Descas}, $\epsilon_c$ ranges
between $0.98$ and $1.01$, i.e., within $\pm 2.5\%$. This relatively small
change leads to significant variation of $\phi$ between $0.5$ and $0.59$. The
same authors have shown that the set of parameters, $\epsilon_c = 1.01$ and
$\phi
= 0.59$, leads to a more accurate scaling prediction. The adsorption of the
tethered SAW chain on a simple cubic lattice for chain lengths of up to $N =
1000$ (by means of the so-called ``scanning method'') gives: $\phi = 0.53 \pm
0.007$ \cite{Meirovitch}. In yet another MC-study, based on the pruned-enriched
Rosenbluth method (PERM) \cite{Grassberger_1}, it was found  that $\phi$ is
pretty close to $0.5$. However, in a more recent study of the same author
\cite{Grassberger_2} one determined for $\phi$  an  even smaller  value: $\phi =
0.484 \pm 0.002$. The value $\phi = 0.5$ was also been supported by the
MC-simulation results based on the off-lattice model \cite{Metzger}.

The analytical methods for calculation of $\phi$ are  based on the
field-theoretical renormalization group (RG) study of the semi-infinite
$n$-vector model in the $n \rightarrow 0$ limit. In earlier investigations
\cite{Diehl_1,Diehl_2,Eisen} the $\varepsilon$-expansion (where $\varepsilon =
4-d$) up to order $\varepsilon^2$ lead to the prediction $\phi = 0.67$ which
deviates widely from all MC-findings. In a more recent investigation the
so-called massive field-theory  approach at fixed $d$ (i.e., the
$\varepsilon$-expansion has been avoided) was extended to systems with surfaces
\cite{Shpot_1,Shpot_2}. The result for the crossover exponent reads $\phi
\approx 0.52$. Thus we believe that the present study elucidates the origin for
the diversity of results concerning the precise value of $\phi$ and provides a
physical background of it.

\section{Adsorption under external detaching force}\label{Force_Induced}
\label{sect_theor_force}
The adsorption of a Gaussian chain on a solid plane under detaching force acting
on the chain end has been studied first by Skvortsov, Gorbunov and Klushin
\cite{Skvortsov_1,Skvortsov_2} in the early 90s. For a Gaussian chain the
problem can be solved rigorously even for a finite chain length $N$. The
adsorption-desorption transition is of the first order, however, phase
coexistence and metastable states are absent.

Below we apply the GC - ensemble approach to the case of self-avoiding polymer
chain adsorption under the presence of detaching force. Again, the problem has
much in common with the unzipping transition of double-stranded DNA
\cite{Kafri}.
When a force $f$ is applied to the free end of the tethered chain, the tail GC
partition function in Eq.(\ref{GC_partition}) changes. The total GC-partition
function is then given by
\begin{eqnarray}
 \Xi (z) = \frac{V_{0}(z) \: R(z)}{1 - V(z) U (z)}
\label{GC_partition_Force}
\end{eqnarray}
where the tail GC - partition function now takes on the form
\begin{eqnarray}
 R(z) = 1 + \sum_{m=1}^{\infty} \: \Xi_{\rm tail} (m) \: z^m
\label{Tail_Force}
\end{eqnarray}
In Eq.(\ref{Tail_Force})  $\Xi_{\rm tail} (m)$ is the canonical partition
function of the tail under applied force:
\begin{eqnarray}
\Xi_{\rm tail} (m) = \frac{\mu_3^m}{m^{\beta}} \: \int d^3 r \: P_m ({\bf r}) \:
\exp \left( f r_{\perp}/T\right).
\label{Canon_Tail_Force}
\end{eqnarray}
Here we take into account that the pulling force is directed perpendicular to
the plane (in $r_{\perp}$-direction). In Eq. (\ref{Canon_Tail_Force}) $P_m ({\bf
r})$ is
the end-to-end distance probability distribution function (PDF).  To estimate
this function on large distances from the solid plane, i.e., at $r_{\perp} \gg
R_m
\approx a m^{\nu}$ (here and in what follows $a$ denotes the length of a
Kuhn-segment), we assume, following Kreer et al. \cite{Binder}, that under
this condition the PDF is given by the  des Cloizeaux expression
\cite{DesCloizeaux} for the bulk:
\begin{eqnarray}
 P_m ({\bf r}) = \frac{1}{R_m^3} \: F\left(\frac{{\bf r}}{R_m}\right)
\label{P}
\end{eqnarray}
where the scaling function $F(x)$ is
\begin{eqnarray}
 F(x) = B x^t \: \exp\left(- D x^{\delta}\right).
\label{F}
\end{eqnarray}
In Eq. (\ref{F}) $B$ and $D$ are constants while the exponents   $\delta$ and
$t$ are given by
\begin{eqnarray}
 \delta = \frac{1}{1 - \nu}
\label{Delta}
\end{eqnarray}
and 
\begin{eqnarray}
 t = \frac{\beta -d/2 + d \nu}{1 - \nu}.
\label{t}
\end{eqnarray}
Here $\beta = 1 - \gamma_1$ is the tail surface exponent and $d = 3$. Note that
in the limit $r_{\perp} \gg R_m$ the only difference between the PDFs in the
bulk and in
the semi-infinite case lies in the fact that instead of the exponent $\gamma$ in
Eq. (\ref{t}) one has $\gamma_1$. The integration over the coordinates
parallel to the plane in Eq.(\ref{Canon_Tail_Force}) is readily carried out and
one obtains
\begin{eqnarray}
\Xi_{\rm tail} (m) &=& \frac{\mu_3^m}{m^{\beta}} \:
\frac{C}{R_m}\int\limits_{0}^{\infty}  d r_{\perp} \:
\left(\frac{r_{\perp}}{R_m}\right)^{2+t-\delta} \: \exp \left[ - D
\left(\frac{r_{\perp}}{R_m}\right)^{\delta} + \frac{f r_{\perp}}{T} \right]
\nonumber\\
&=& C \: \frac{\mu_3^m}{m^{\beta}} \: \int\limits_{0}^{\infty}  d x \: x^{2 + t
- \delta} \exp \left(- D x^{\delta} + {\tilde f}_m x \right)
\label{Canon_Tail_Force_1}
\end{eqnarray}
where the normalization constant $C = \delta D^{(3+t)/\delta -
1}/\Gamma[(3+t)/\delta - 1]$. The integral in Eq. (\ref{Canon_Tail_Force_1}) can
be tackled by the saddle point method (since ${\tilde f}_m \equiv f R_m/T \gg
1$). The saddle point itself is defined by the value $x_{\rm sp} = ({\tilde
f}_m/(\delta D))^{1/(\delta - 1)} \sim {\tilde f}_m^{1/\nu - 1}$, or, in terms
of the $r_{\perp}$-variable,
\begin{eqnarray}
 r_{\perp}^{\rm sp} \approx R_m \: \left({\tilde f}_m \right)^{1/\nu -
1}\approx a m
\left(\frac{f a}{T}\right)^{1/\nu - 1}
\label{Pincus}
\end{eqnarray}
which is nothing but the well-known Pincus deformation law \cite{Gennes}.
Finally, Eq. (\ref{Canon_Tail_Force_1}) becomes
\begin{eqnarray}
\Xi_{\rm tail} (m) = a_1 \: ({\tilde f})^{\theta} \: \frac{\mu_3^m}{m^{\beta -
\theta \nu}} \: \exp \left(a_2 {\tilde f}^{1/\nu} m\right)
\end{eqnarray}
with $a_1$ and $a_2$ being constants, the dimensionless force ${\tilde f} \equiv
f a/T$, and the exponent $\theta = (2+t -3\delta/2)/(\delta - 1)$. Thus the
GC-partition function , Eq. (\ref{Tail_Force}),  can be written as
\begin{eqnarray}
R (z) &=& 1 + a_1 \:{\tilde f}^{\theta} \:  \sum_{m=1}^{\infty} \:
\frac{1}{m^{\psi}} \: \left[z \mu_3 \exp (a_2 {\tilde
f}^{1/\nu})\right]^m\nonumber\\
&=& 1 + a_1 \:{\tilde f}^{\theta} \: \Phi (\psi, z \mu_3 \exp (a_2 {\tilde
f}^{1/\nu}))
\label{GC_Tail_Force}
\end{eqnarray}
where we have defined the new exponent
\begin{eqnarray}
 \psi = \beta - \nu \theta = \frac{d-1}{2} - (d-2) \nu .
\end{eqnarray}
One should point out that the exponent $\beta$ drops out from the final
expression for $\psi$ which for $d=3$ is defined as $\psi = 1 - \nu$.

It is evident from Eq. (\ref{GC_Tail_Force}) that (cf. Eq.(\ref{Cases}))
at $z \rightarrow \mu_3^{-1} \: \exp (- a_2 {\tilde f}^{1/\nu})$ the tail
GC-partition function has a branch point at $z = z^{\#}$, i.e.
\begin{eqnarray}
R(z) \sim a_1 \: ({\tilde f})^{\theta} \: \frac{ \Gamma (1 - \psi)
(z^{\#})^{1-\psi}}{\left[z^{\#} - z\right]^{1-\psi}}
\label{Branch_Singularity}
\end{eqnarray}
where $1 - \psi < 1$ and  
\begin{eqnarray}
 z^{\#} = \mu_3^{-1} \: \exp (- a_2 {\tilde f}^{1/\nu}) .
\label{Branch_Value}
\end{eqnarray}

Turning back to the total GC-partition function, Eq.(\ref{GC_partition_Force}),
one may conclude that $\Xi (z)$ has two singularities on the real axis ${\rm
Re} z$:  the pole $z^{*}$ which is defined by Eq.(\ref{Pole}), and  the branch
point  $z^{\#}$ given by Eq. (\ref{Branch_Value}). It is well known (see, e.g.,
Sec. 2.4.3 in \cite{Rudnick}) that in the thermodynamic limit, $N\rightarrow
\infty$, the contribution to the coefficient of  $z^{N}$ (i.e., to $\Xi_N$)
consists of contributions by the pole and by the branch singular points, i.e.
\begin{eqnarray}
 \Xi_N \sim C_1 \: (z^{*})^{-(N+1)} + \frac{C_2}{\Gamma(1-\psi)} \: N^{-\psi} \:
(z^{\#})^{-(N+1-\psi)}
\label{Pole_Branch}
\end{eqnarray}
The singular points, $z^{*}$ and $z^{\#}$, are involved in
Eq.(\ref{Pole_Branch}) with large negative exponents. Hence, for large $N$ only
the smallest of these points matters. On the other hand, $z^{*}$ depends on the
dimensionless adsorption energy $\epsilon$ only (or, on $w=\exp(\epsilon)$)
whereas $z^{\#}$ is controlled by the dimensionless external force ${\tilde f}$
(cf., Eq.(\ref{Branch_Value})). Therefore, in terms of the two {\it control
parameters}, $\epsilon$ and ${\tilde f}$, the equation
\begin{eqnarray}
 z^{*}(\epsilon) = z^{\#}({\tilde f})
\label{Detachment_Line}
\end{eqnarray}
determines the critical line of transition between the adsorbed phase and the
force-induced desorbed phase. In the following this line will be referred to as
the {\it detachment line}. The controll parameters, $\epsilon_D$ and
${\tilde f}_D$, which satisfy Eq. (\ref{Detachment_Line}), will be named
detachment
energy and detachment force, respectively. On the detachment line the system
undergoes a first-order phase transition. The detachment line at ${\tilde f}_D
\rightarrow 0$ terminates in the critical adsorption point, $\epsilon_c$, where
the transition becomes of second order. In the vicinity of the critical
adsorption point the detachment force ${\tilde f}_D$ behaves as
\begin{eqnarray}
 {\tilde f}_D \sim (\epsilon - \epsilon_c)^{\nu/\phi}
\label{Detachment_Force} 
\end{eqnarray}
where we have used Eq.(\ref{Detachment_Line}) as well as  Eqs.
(\ref{Solution}) and (\ref{Branch_Value}).

\subsection{Order parameter}\label{OP}

Let us study first how the fraction of adsorbed monomers $n = N_s/N$, which we
use as an order parameter, depends on the pulling force at fixed value of
the contact energy $\epsilon_1 > \epsilon_c$. For ${\tilde f} < {\tilde f}_D$
it is clear that $z^{*} < z^{\#}$ and the first term in Eq. (\ref{Pole_Branch})
dominates over the second one. In this case the order parameter
\begin{eqnarray}
 n = - \left. \frac{\partial \ln z^{*}(w)}{\partial \ln
w}\right|_{w=\exp(\epsilon_1)}
\end{eqnarray}
is  constant independent of the  force. At ${\tilde f} >  {\tilde f}_D$  (i.e.,
after crossing the detachment line) $z^{*} > z^{\#}$ and the second term in Eq.
(\ref{Pole_Branch}) prevails. Since  $z^{\#}$ is $w$-independent, it is evident
that $n = 0$, i.e., the polymer is totally detached. In result, the $n$ vs.
${\tilde f}$ dependence resembles a step - function with a jump at ${\tilde f} =
{\tilde f}_D$.

Now let us fix the force ${\tilde f} = {\tilde f}_1$ and investigate how the
order parameter $n$  depends on the adsorption energy $\epsilon$ or on the
fugacity $w$. Again, Eq.(\ref{Detachment_Line}) at ${\tilde f} = {\tilde f}_1$
defines a detachment energy $\epsilon_D$.  At $\epsilon < \epsilon_D$ one has
still $z^{\#} < z^{*}$ and the second term in Eq. (\ref{Pole_Branch})
dominates 
so that the chain is completely desorbed (i.e., $n = 0$). At $\epsilon >
\epsilon_D$ only the first term in Eq.(\ref{Pole_Branch}) survives so that the
relationship $n$ vs. $\epsilon$ follows the conventional adsorption dependence
without any force-influence. The transition at $\epsilon = \epsilon_D$  is of
first order whereby the order parameter jump grows as the force  ${\tilde f}_1$
increases. 
 
\begin{figure}[htb]
\vspace{0.8cm}
\includegraphics[scale=0.4,angle=-90]{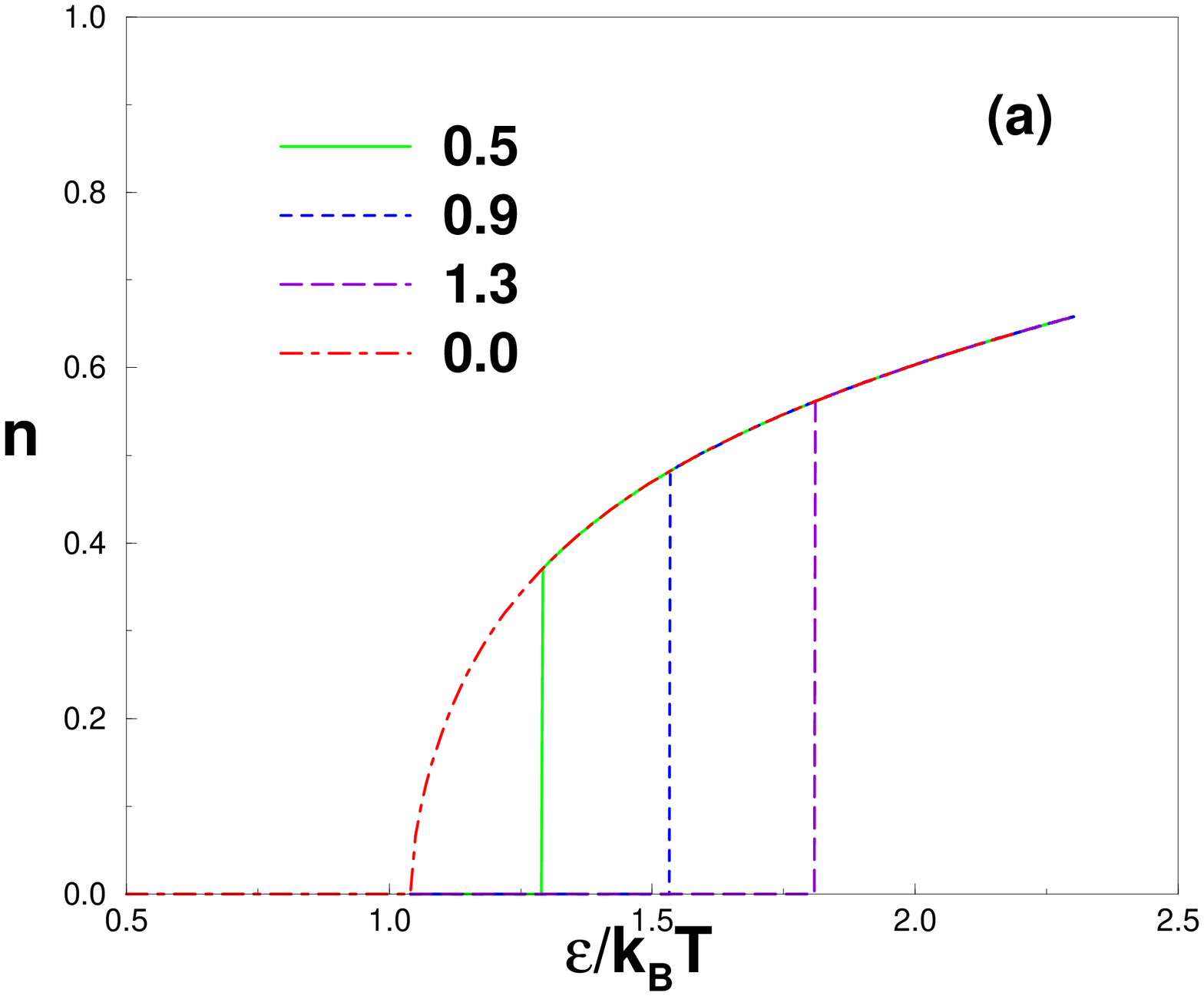}
\includegraphics[scale=0.4,angle=-90]{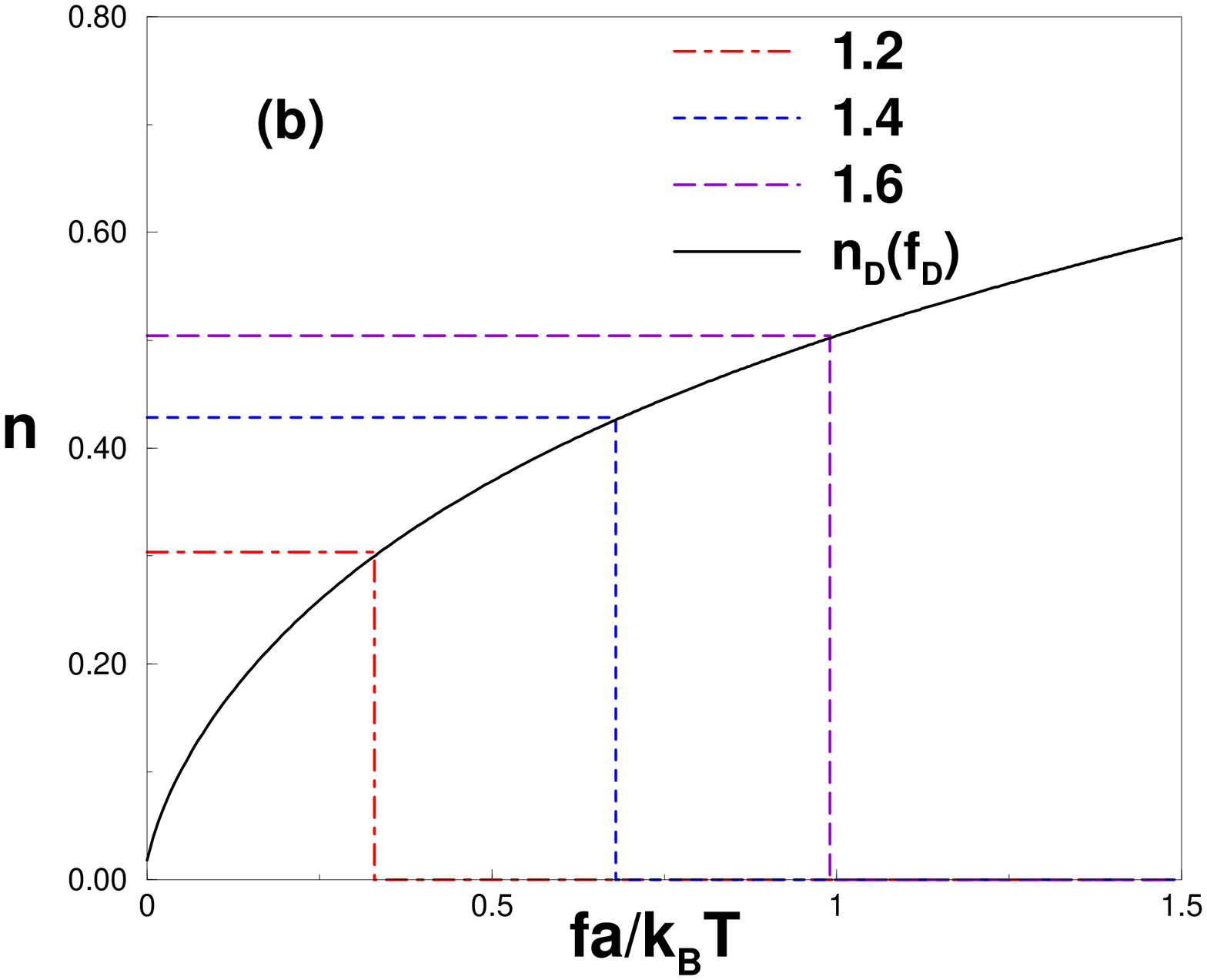}
\includegraphics[scale=0.4,angle=-90]{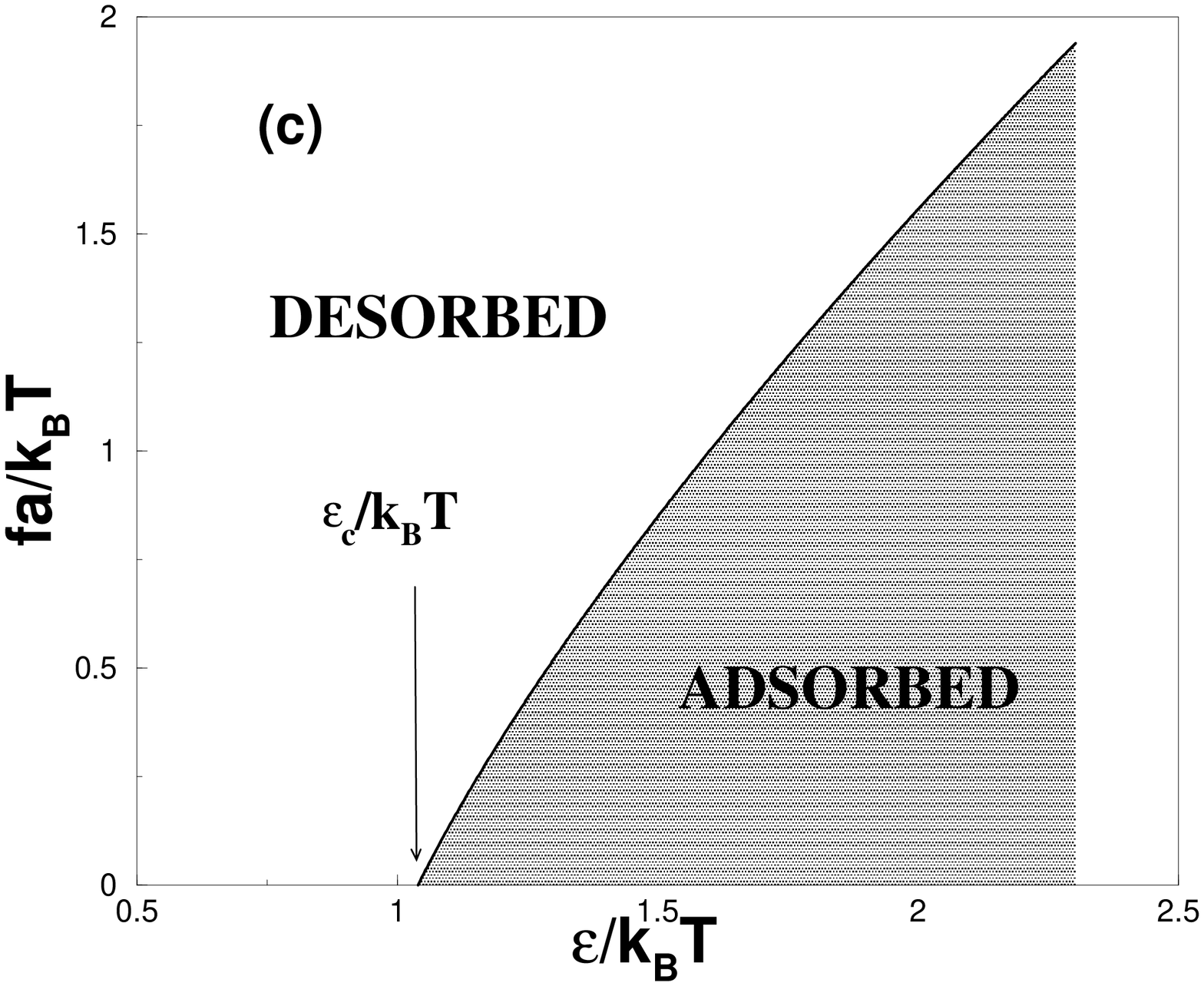}
\includegraphics[scale=0.4,angle=-90]{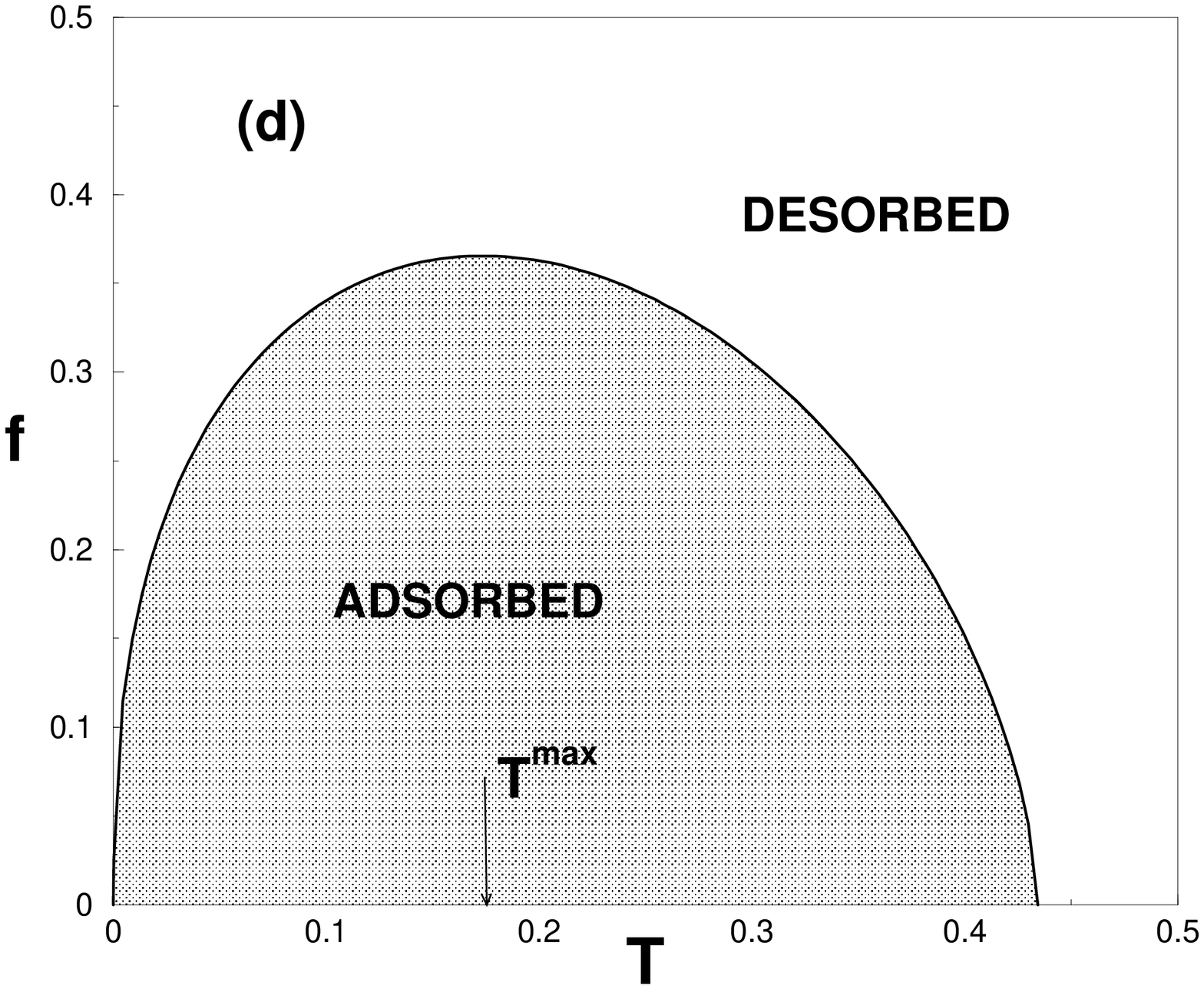}
\caption{The 'order parameter' , $n$, against the: (a) energy of adsorption
$\epsilon$ ($f$ is given as a parameter); (b) - against the pulling force $f$,
with $\epsilon$ a parameter. Vertical lines denote the discontinuous
jumps of $n$, indicating a $1^{st}-$ order transition. The $n_D(f_D)$ (full
line) in (b) denotes the order parameter value at the detachment line. (c) The
phase diagram of the adsorption-desorption transition under puling force $f$ in
dimensionless units. An arrow at $\epsilon_c$ denotes the point of critical
adsorption for $f = 0$. (d) The {\em reentrant} phase diagram - the same as in
(c) but with force against temperature in dimensional units at fixed value of
$\epsilon > \epsilon_c$. The largest force $f$ for which chain adsorption may
still take place occurs at temperature $T^{max}$, as indicated by an arrow.
}
\label{Detachment_theor}
\end{figure} 
In Figure \ref{Detachment_theor}a,b we show the predicted variation of the
order parameter for an infinitely long chain, following from the present
consideration. The boundary of the region of adsorbtion, shown in the phase
diagram in Figure \ref{Detachment_theor}c, denotes the line of critical values
of detachment force for any given attraction of the substrate as described by
Eq. (\ref{Detachment_Line}).
 
The adsorption-desorption first order phase  transition under pulling force has
a clear dichotomic nature (i.e., it follows an ``either - or'' scenario): in the
thermodynamic limit $N \rightarrow \infty$ there is {\em no} phase coexistence!
The configurations are divided into adsorbed and detached (or stretched)
dichotomic classes. The metastable states are completely absent.  Basically,
this is in line with the general thermodynamic principles which argue that in
thermal equilibrium the thermodynamic potentials are convex functions of their
order parameters. This exclude multiple minima and metastable states
\cite{Binder_Review}.
\subsection{Reentrant behavior of the phase diagram}

The results given in Section \ref{OP} demonstrate that the detachment line on
the phase diagram is a monotonous function in terms of the dimensionless
quantities ${\tilde f}_D$ vs. $\epsilon_D$. Recently, it has been revealed that
the detachment line, when represented in terms of {\em dimensional} variables,
force $f_D$ versus temperature $T$, goes (at the relatively low temperature)
through a maximum, that is, the desorption transition shows a reentrant
behavior! Below we demonstrate that this result follows directly from our
theory.

First, one should note that the low temperature limit implies large values of
the ratio $\epsilon = \varepsilon/k_BT$. On the other hand, the solution
$z^{*}(w)$, which results from Eq.(\ref{Basic_Eq}), goes to zero, i.e., $z^{*}
\rightarrow 0$, when $\varepsilon \rightarrow \infty$. One may assume that under
these conditions $z^{*} \mu_2 {\rm e}^{\epsilon} \rightarrow 1^{-}$ (this will
be proven {\it a posteriori}). Then the polylog function in the l.h.s. of
Eq.(\ref{Basic_Eq}) reads $\Phi (\alpha, \mu_3 z^{*}) \approx \mu_3 z^{*}$ but
$\Phi^{-1} (\lambda, \mu_2 w  z^{*}) \approx c_1 (1 - \mu_2 w
z^{*})^{1-\lambda}$ (where we have used Eq. (\ref{Cases}) and the fact that
$\lambda < 1$). Taking into account Eq.(\ref{Basic_Eq}), one arrives at the
following result
\begin{eqnarray}
 \mu_3 z^{*} \approx c_1 (1 - \mu_2 w  z^{*})^{1-\lambda}
\label{Z_star}
\end{eqnarray}
This equation determines the function $z^{*}(w)$ at large $w$. To zero-order
approximation the solution reads $z^{*}_{(0)} \approx (\mu_2 w)^{-1}$. Within
the first order approximation $z^{*}_{(1)} \approx (\mu_2 w)^{-1} - \delta$
where the decrement $\delta$ is found as $\delta = (1/\mu_2 w) (\mu_3/\mu_2
w)^{1/(1-\lambda)}$. This result is consistent with the assumption $z^{*} \mu_2
{\rm e}^{\epsilon} \rightarrow 1^{-}$ so that the solution of Eq. (\ref{Z_star})
in the main approximation can be written as
\begin{eqnarray}
 z^{*} \approx \frac{1}{\mu_2} \: {\rm e}^{-\epsilon}
\label{Large_Epsilon}
\end{eqnarray}
By making use of this solution as well as of the result given by
Eq.(\ref{Branch_Value}) in the Eq.(\ref{Detachment_Line}), the detachment  line
at large dimensionless detachment energy $\epsilon_D \equiv \varepsilon/T$ and
force ${\tilde f}_D \equiv a f_D/T$ can be written as
\begin{eqnarray}
{\tilde f}_D = \frac{1}{a_2^{\nu}} \: \left[\epsilon_D  - \ln \left(
\frac{\mu_3}{\mu_2}\right) \right]^{\nu}. 
\label{Force_vs_Energy}
\end{eqnarray}
Thus, in terms of the dimensionless controll parameters ${\tilde f}_D$ increases
as the energy  $\epsilon_D$ increases. Notably, however, if the same detachment
line is represented in terms of the dimensional control parameters, detachment
force $f_D$ vs. detachment temperature $T_D$ (with the dimensional adsorption
energy $\varepsilon_0$ being fixed), one encounters a nonmonotonic behavior
\begin{eqnarray}
f_D = \frac{T_D}{a} \: \left[ \frac{\varepsilon_0}{T_D} - \ln \left(
\frac{\mu_3}{\mu_2}\right)\right]^{\nu}
\label{Force_vs_Temperature}
\end{eqnarray}
which is shown in Fig.~\ref{Detachment_theor}d. The curve given by
Eq.(\ref{Force_vs_Temperature}) goes through a maximum at a temperature given by
\begin{eqnarray}
 T_D^{\rm max} = \frac{(1 - \nu) \varepsilon_0}{\ln
\left(\frac{\mu_3}{\mu_2}\right)}.
\label{Max}
\end{eqnarray}
 Such nonmonotonic behavior is termed {\it reentrant} and can be observed in the
DNA unzipping process \cite{Maritan,Seno,Shakhnovich} as  well as in the case
of 
stretched polymer adsorption on solid surfaces \cite{Mishra,Krawczyk}. At very low
$T$, however, the expression, Eq.~(\ref{P}), for $P_m({\bf r})$ \cite{DesCloizeaux}
predicts divergent chain deformation \cite{Seno}, i.e., it becomes
unphysical. One can readily show that in this case the correct behavior is
given by $f a = \varepsilon_0 + T\ln(\mu_3/\mu_2)$.

\subsection{Average loop and tail lengths close to the detachment line}

\begin{figure}[htb] 
\includegraphics[scale=0.4,angle=-90]{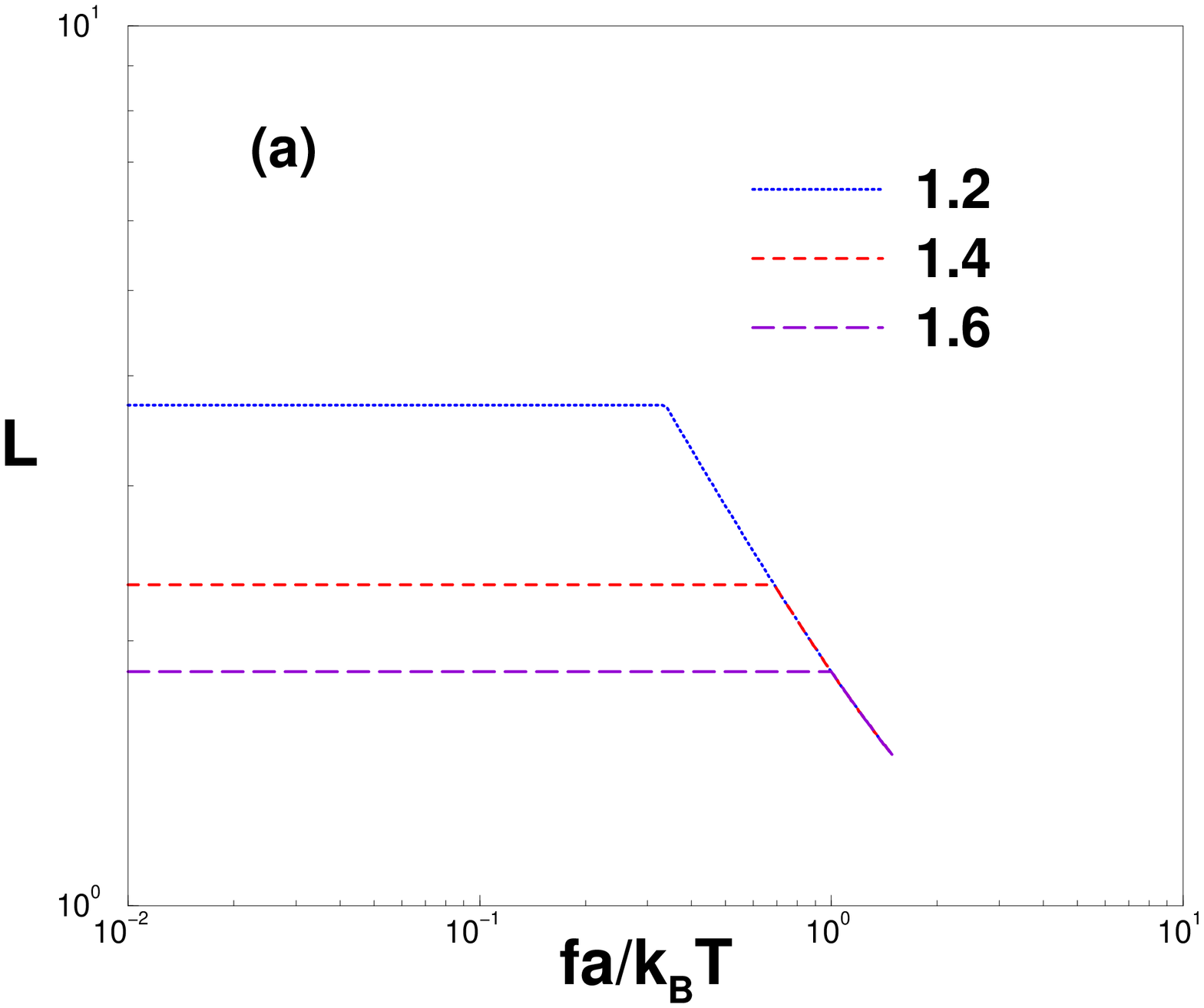}
\includegraphics[scale=0.4,angle=-90]{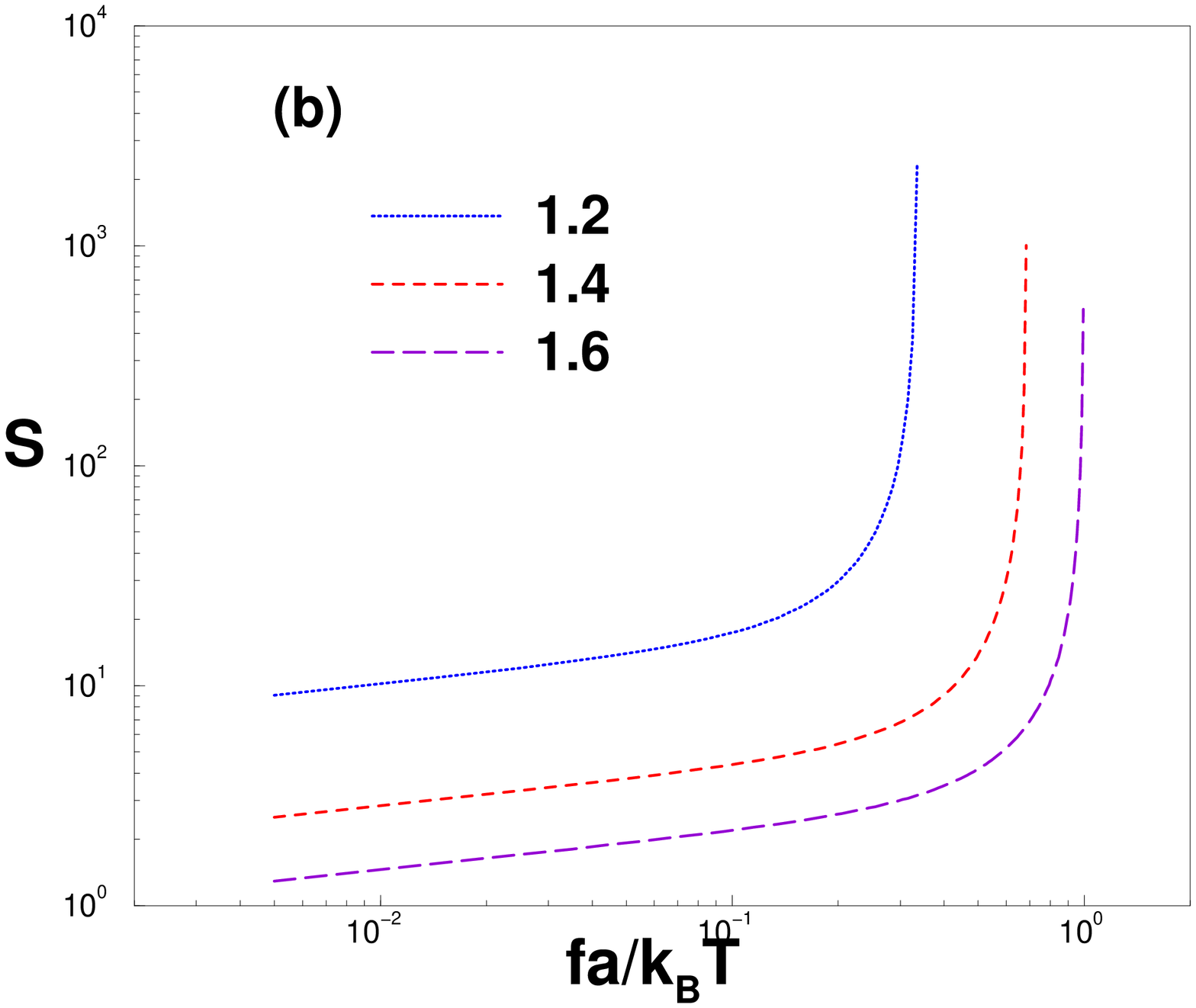}
\caption{Variation of the average loop size, $L$, with detachment force
strength $f$  for several values of the adsorption energy $\epsilon$ (given as
a parameter). (b) Mean tail size $S$ against $f$ at different
substrate attraction $\epsilon$.
}
\label{L_S_f}
\end{figure}

As long as the adsorption energy $\epsilon > \epsilon_c$ (or $w > w_c$), the
average loop length $L$ remains finite upon the detachment line crossing 
Namely, at ${\tilde f} < {\tilde f}_D$ the fugacity $z = z^{*}(w)$ and the
average loop length are given by
\begin{eqnarray}
 L = \left. z \: \frac{\partial \ln U(z)}{\partial z}
\right|_{z=z^{*}(w)} = \frac{\Phi \left(\alpha - 1,
\mu_3 z^{*}(w)\right)}{ \Phi \left(\alpha, \mu_3 z^{*}(w)\right)}.
\label{Loop_Length_Detachment_1}
\end{eqnarray}
Thus, at ${\tilde f} < {\tilde f}_D$ the force does not effects the loop length.
At ${\tilde f} > {\tilde f}_D$ the fugacity is given by $z = z^{\#}({\tilde f})$
where $z^{\#}$ is determined from Eq. (\ref{Branch_Value}). In this case the
average loop length reads
\begin{eqnarray}
 L = \left. z \: \frac{\partial \ln U(z)}{\partial z} \right|_{z=z^{\#}
({\tilde f})} = \frac{\Phi \left(\alpha - 1, \mu_3 z^{\#}({\tilde f})\right)}{
\Phi \left(\alpha, \mu_3 z^{\#}({\tilde f})\right)}
\label{Loop_Length_Detachment_2}
\end{eqnarray}
Recall, that at $\epsilon > \epsilon_c$ and ${\tilde f} > {\tilde f}_D$  we have
$\mu_3 z^{\#} < \mu_3 z^{*} <1$. In this case the function  given by Eq.
(\ref{Loop_Length_Detachment_2}) declines when the force grows - see Figure
\ref{L_S_f}a.

In contrast, the average tail length $S$ diverges in the vicinity of the
detachment line. Indeed, at ${\tilde f} < {\tilde f}_D$ the average tail length
is given by
\begin{eqnarray}
 S = \left. z \: \frac{\partial \ln R(z)}{\partial z} \right|_{z=z^{*}(w)}  &=&
\frac{a_1 {\tilde f}^{\theta}\Phi \left(\psi - 1, \mu_3 z^{*}(w)\exp(a_2 {\tilde
f}^{1/\mu})\right)}{1 + a_1 {\tilde f}^{\theta} \Phi \left(\psi, \mu_3 z^{*}(w)
\exp(a_2 {\tilde f}^{1/\mu})\right)}\nonumber\\
&\approx& \frac{1}{\left[ 1 - \mu_3 z^{*} \exp (a_2 {\tilde f}^{1/\mu})\right] }
\label{Tail_Length_Detachment}
\end{eqnarray}
because $\psi < 1$ and $\mu_3 z^{*} \exp (a_2 {\tilde f}^{1/\mu}) \leq 1$ (cf. 
Eq. (\ref{Cases})). In the vicinity of the detachment line $1 - \mu_3
z^{*}(w)\exp(a_2 {\tilde f}^{1/\mu}) \approx ({\tilde f}_D - {\tilde f})/(\nu
{\tilde f}_D)$ and, therefore, 
\begin{eqnarray}
 S \propto \frac{{\tilde f}_D}{{\tilde f}_D - {\tilde f}}.
\label{Tail_Length_Divergency}
\end{eqnarray}
At ${\tilde f} \ge {\tilde f}_D$ the fugacity $z=z^{\#}({\tilde f})$ and hence,
\begin{eqnarray}
 S = \left. z \: \frac{\partial \ln R(z)}{\partial z} \right|_{z=z^{\#}
({\tilde f})}  \rightarrow \infty .
\label{Very_Long_Tail}
\end{eqnarray}
The divergence in Eq.(\ref{Very_Long_Tail}) follows immediately from
Eq.(\ref{Branch_Singularity}) which holds in the thermodynamical limit. In
practice, however, for a large but finite chain length $S \rightarrow N$
at ${\tilde f} \geq {\tilde f}_D$. Thus, despite the abrupt first order phase
transition, as far as the order parameter $n$ is concerned, the detachment in
terms of the tail length $S$ starts diverging already at ${\tilde f} \leq
{\tilde f}_D$ as one comes close to the critical detachment force ${\tilde
f}_D$.

\subsection{Latent heat variation upon detachment}
 
What is the internal energy change while crossing the detachment line?  At
${\tilde f} > {\tilde f}_D$ the stretching energy  $E$ follows the Pincus law,
so that
\begin{eqnarray}
 E ({\tilde f} = {\tilde f}_D + 0) = N T {\tilde f}_D^{1/\nu}
\label{Pincus_Energy}
\end{eqnarray}
In the adsorbed phase 
\begin{eqnarray}
 E ({\tilde f} = {\tilde f}_D - 0) = - N T \: \epsilon_D \: n(w_D)
\label{Adsorbed_Energy}
\end{eqnarray}

In result, the latent heat $q$, consumed upon  detachment (or,  due to
force-induced desorption,) reads
\begin{eqnarray}
 q \equiv E ({\tilde f} = {\tilde f}_D + 0) - E ({\tilde f} = {\tilde f}_D - 0)
= NT \left[ {\tilde f}_D^{1/\nu} + \epsilon_D \: n(w_D)\right]  > 0
\label{Latent_Heat}
\end{eqnarray}
i.e., the heat is absorbed by the system during the force -induced desorption.
In the vicinity of the critical point ${\tilde f}_D \sim (\epsilon -
\epsilon_c)^{\nu/\phi}$ and $n \sim (\epsilon - \epsilon_c)^{1/\phi - 1}$, thus
to
a leading order
\begin{eqnarray}
 q \approx N T \epsilon_c \:  (\epsilon - \epsilon_c)^{1/\phi - 1}.
\end{eqnarray}
\section{Monte Carlo Simulation Model}
\label{sect_MC_model}
We have investigated the force induced desorption of a polymer by means of
extensive Monte Carlo simulations. We use a coarse grained off-lattice
bead-spring model \cite{MC_Milchev} which has proved rather efficient in a
number of polymers studies so far. The system consists of a single polymer chain
tethered at one end to a flat impenetrable structureless surface. The
effective bonded interaction is described by the FENE (finitely extensible
nonlinear elastic) potential:
\begin{equation}
U_{FENE}= -K(1-l_0)^2ln\left[1-\left(\frac{l-l_0}{l_{max}-l_0} \right)^2 \right]
\label{fene}
\end{equation}
with $K=20, l_{max}=1, l_0 =0.7, l_{min} = 0.4$. In fact, $l_{max}$ sets the
length scale in our model. The nonbonded interactions
between monomers are described by the Morse potential.
\begin{equation}
\frac{U_M(r)}{\epsilon_M} =\exp(-2\alpha(r-r_{min}))-2\exp(-\alpha(r-r_{min}))
\end{equation}
with $\alpha =24,\; r_{min}=0.8,\; \epsilon_M/k_BT=1$.

The surface interaction is described by a square well potential,
\begin{equation}\label{surf_pot}
 U_w(z) = \begin{cases}
\epsilon, & z < \delta \\
0, & z \ge \delta 
          \end{cases}
\end{equation}
where the range of interaction $\delta = l_{max} / 4$. The strength $\epsilon$
of the surface potential is varied from $2.0$ to $7.0$ and $k_BT=1$.

We employ periodic boundary conditions in the $x-y$ directions and impenetrable
walls in the $z$ direction. The lengths of the studied polymer chains are
typically $32$, $64$, and  $128$. The size of the simulation box was chosen
appropriately to the chain length, so for example, for a chain length of $128$,
the box size was $256 \times 256 \times 256$ . All simulations were carried out
for constant force, that is, in the stress ensemble. A force $f$ was applied to
the last  monomer in the $z$-direction, i.e., perpendicular to the
adsorbing surface.

The standard Metropolis algorithm was employed to govern the moves with  self
avoidance automatically incorporated in the potentials. In each Monte Carlo
update, a monomer was chosen at random and a random displacement attempted with
$\Delta x,\;\Delta y,\;\Delta z$ chosen uniformly from the interval $-0.5\le
\Delta x,\Delta y,\Delta z\le 0.5$. If the last monomer was displaced in $z$
direction, there was an energy cost of $-f\Delta z$ due to the pulling force.
The transition probability for the attempted move was calculated from the change
$\Delta U$ of the potential energies before and after the move was performed as
$W=exp(-\Delta U/k_BT)$. As in a standard Metropolis algorithm, the attempted
move was accepted, if $W$ exceeds a random number uniformly distributed in the
interval $[0,1]$.

As a rule, the polymer chains have been originally equilibrated in the MC method
for a period of about $ 5 \times 10^5$ MCS after which typically $500$
measurement runs were performed, each of length $2 \times 10^6$ MCS. The
equilibration period and the length of the run were chosen according to the
chain length and the values provided here are for the longest chain length.

\section{Monte Carlo Simulation Results}
\label{sect_results}
\subsection{Determination of the  detachment point}

In the absence of external pulling force, the transition of a polymer from
desorbed to adsorbed state is known to be of second order, and the
fraction of adsorbed  monomers, $n$, can be identified as an order parameter.
Therefore, in our computer experiment we use $n$ to determine the point of
polymer detachment from the adsorbing surface.
\begin{figure}[bht]
\vspace{0.8cm}
\includegraphics[scale=0.4]{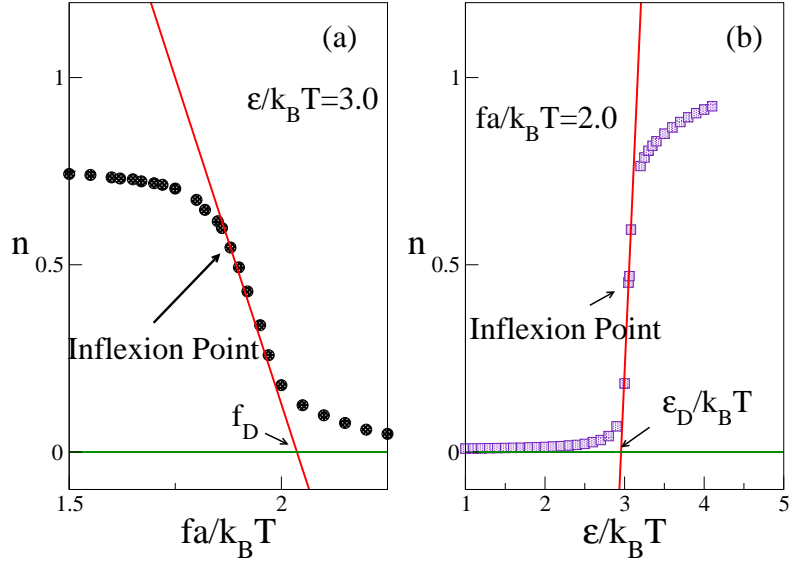}
\caption{(a) Plot of the 'order parameter', $n$, against pulling force $f$, for
an adsorption strength $\varepsilon/k_BT=3.0$. The polymer chain length is
$N$=128. The tangent at the inflexion point of the curve meets the abscissa at
$f_D$ which we define as the detachment force. (b) The 'order parameter', $n$,
against the adsorption potential $\epsilon$ for fixed pulling force $f=2.0$.
The tangent at the inflexion point of the curve meets the abscissa at
$\varepsilon_D/k_BT$ where the polymer adsorbs at the surface plane.
}
\label{Detachment}
\end{figure}
At constant surface potential, $\epsilon$, one finds that $n$ steeply decreases
upon a small increase of the pulling force whereby the polymer chain undergoes a
transition from an adsorbed phase  to a grafted-detached state. In order to
locate the point of chain detachment, we draw a tangent at the inflexion point
of the curve $n$ vs. $f$. The detachment force, $f_D$, is then identified as the
point where the tangent intersects the  abscissa ($f$-axis) - see
Fig.~\ref{Detachment}(a). Thus one can determine the  detachment force as a
function of the adsorption potential $\epsilon$. Alternately, from the plot of
$n$ against the adsorption potential $\epsilon$, with the pulling force $f$ held
constant, one can observe that as sharp growth of $n$ as the potential is
slightly increased. The critical potential for chain attachment at the
transition point can be found similarly as indicated in
Fig.~\ref{Detachment}(b).

\begin{figure}[bht]
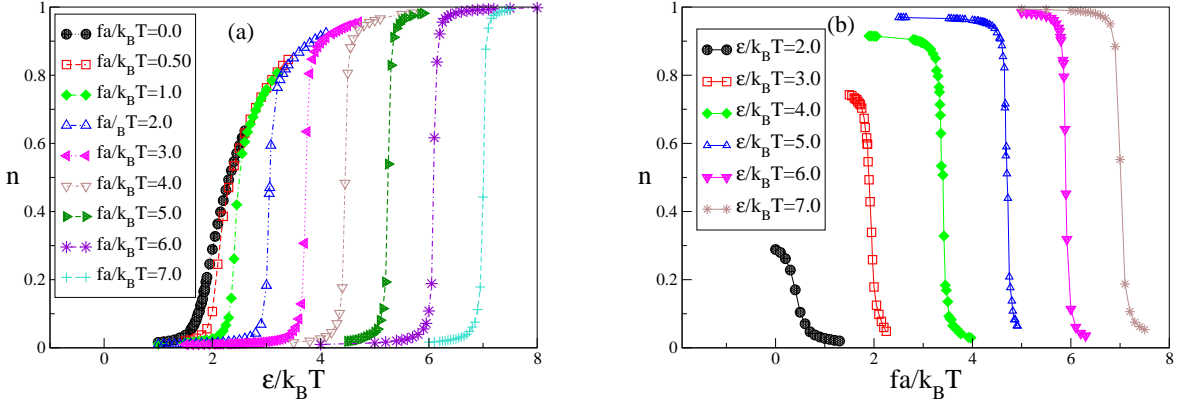

\vspace{0.8cm}
\includegraphics[scale=0.3]{fig5.eps}
\hspace{1cm}
\includegraphics[scale=0.3]{fig4.eps}
\caption{(a) The 'order parameter', $n$, against the surface potential,
$\epsilon$, for various pulling forces. The chain has length $N$=128. (b)
Variation of $n$ with the pulling force, $f$, for several surface potentials.
}
\label{Order-Parameter}
\end{figure}

Fig.~\ref{Order-Parameter}(a) shows the variation of the order parameter with
changing surface potential for several values of the pulling force. Evidently,
the larger the pulling force, the stronger the surface potential, needed to
keep the polymer adsorbed on the plane. In the absence of a force, the order
paramenter changes smoothly. For larger forces, however, the transition becomes
rapidly abrupt. This abrupt behavior of the order parameter is in close
agreement with our theoretical predictions, depicted in
Fig.~\ref{Detachment_theor}. In Fig. ~\ref{Order-Parameter}(b) we show the
variation of the order parameter $n$ with changing force $f$ for various
adsorption potentials $\epsilon$. The threshold values for polymer desorption,
$\epsilon_D(f)$ and $f_D(\epsilon)$, as obtained for chains of different length,
are then extrapolated to obtain the corresponding values  in the thermodynamic
limit $ \rightarrow \infty$.

Our observations show that $\epsilon_D$ increases slightly
(i.e., the finite-size effects are rather small) with growing chain length $N$.
By extrapolating the data to $1/N\rightarrow 0$ one obtains then $\epsilon_D$
for infinite length of the polymer chain. Similarly, the detachment force at
fixed  surface potential $\epsilon$ may be determined in the thermodynamic
limit.

\subsection{Adsorption-desorption phase diagram under pulling}

\begin{figure}[bht]
\vspace{0.8cm}
\includegraphics[scale=0.35]{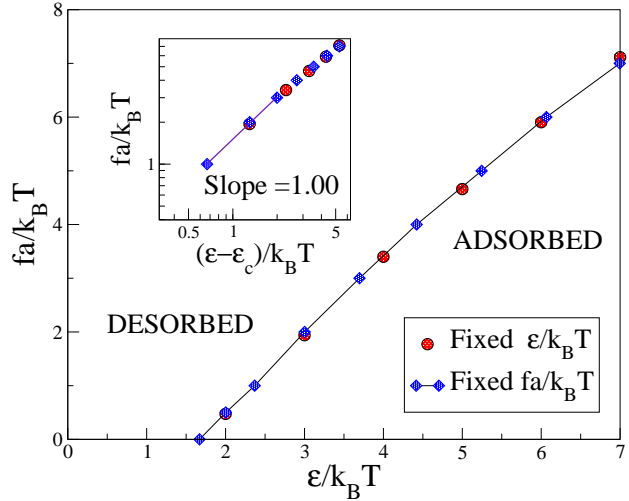}
\caption{Plot of the critical detachment force $f_D$ against the surface
potential $\epsilon$. In the inset in a double logarithmic plot  $f_D$ is
plotted against $(\epsilon-\epsilon_c)/k_BT$. The critical adsorption potential
for zero force has been found earlier \cite{Bhatt_stat_ads} to be
$\epsilon_c=1.67$. }
\label{Phase-diagram}
\end{figure}

Using the threshold values of $f_D$ and $\epsilon_D$ for critical
adsorption/detachment in the thermodynamic limit, one  can construct the
adsorption-desorption phase diagram for a polymer chain. The phase diagram may
be obtained by any of the two methods, i.e., (i) by fixing of the force and
locating $\epsilon_D$, and/or (ii), by fixing of the surface potential and
locating the detachment force $f_D$). The resulting phase diagram is displayed
in Figure \ref{Phase-diagram}. The inset in Fig~\ref{Phase-diagram} shows that
$f_D \propto (\epsilon -\epsilon_c)^{0.97}$ which may be compared to the
theoretical prediction $f_D\sim (\epsilon -\epsilon_c)^{\nu/\phi}$. Hence, this
method gives us an estimate for the crossover exponent $\phi$. For
$\epsilon_c=1.67$ , we find $\phi\sim0.59 \pm0.02$.

\subsection{Average lengths of loops and tails}
\begin{figure}[bht]
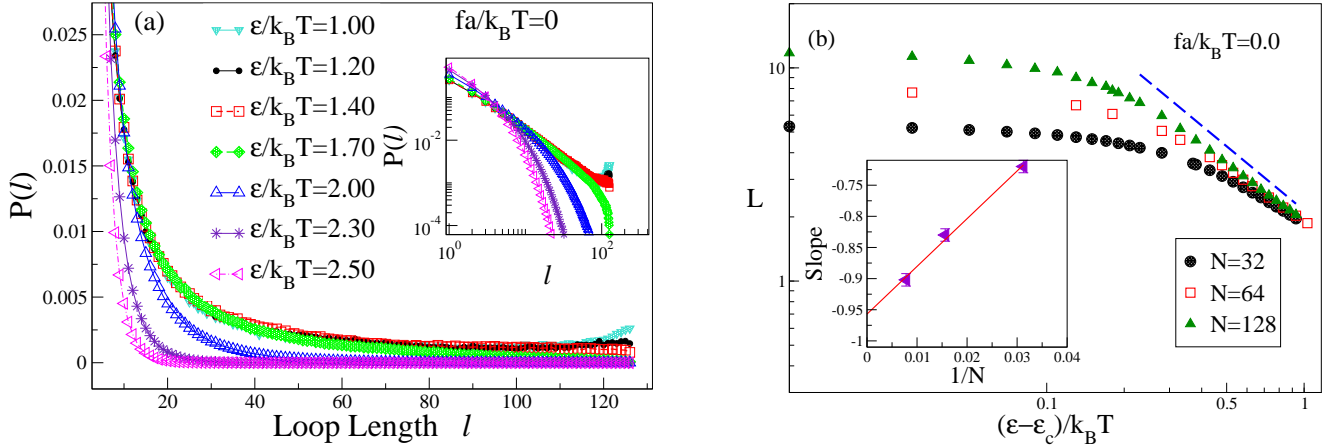

\vspace{1.0cm}
\includegraphics[scale=0.35]{fig7.eps}
\hspace{1cm}
\vspace{0.5cm}
\includegraphics[scale=0.3]{fig12.eps}
\caption{(a) Distribution of loop sizes for chain length $N=128$ at different
strength of the surface potential and no pulling force. The inset shows the same
in double logarithmic coordinates. The measured slope at $\epsilon_c/k_BT =
1.67$ (full line) is $ -1.38 \pm 0.02 $ which practically coincides with the
prediction Eq.(\ref{Loop_Distr}). (b) The average loop length plotted against
$(\epsilon -\epsilon_c)/k_BT$ where $\epsilon_c/k_BT = 1.67$, for various chain
lengths in double logarithmic coordinates. The slopes $x$, indicated by a dashed
line, are obtained from the $L$ vs. $(\epsilon -\epsilon_c)/k_BT$ curves, and
plotted against $1/N$ in the inset. Extrapolation to $1/N \rightarrow 0$ yields
$x \approx 0.95$. }
\label{loop_f0}
\end{figure}
In Fig~\ref{loop_f0}a we plot the PDF of the loop sizes for a chain with $N=128$
at several strengths of the adsorption potential $\epsilon$ in the absence of
pulling. One may readily verify that the PDF has a peak for loops of size unity
which suggests that most frequently single-segment defects (that is, vacancies
in the monomer trains) occur in the conformation of adsorbed chain. However,
for $\epsilon < \epsilon_c$ one may detect clearly in Fig~\ref{loop_f0}a slight
increase in the distribution for loops of size $l \approx N$ which becomes more
pronounced at smaller $\epsilon \approx 1.0 \div 1.2$ in full agreement with
the double-peaked shape, predicted by Eq.~(\ref{Loop_distribution_all}).  

The average
loop size $L$ is plotted against the surface potential (with regard to its
critical value at the adsorption point), $(\epsilon -\epsilon_c)/k_BT$ in
Fig~\ref{loop_f0}b. We find that, well inside the region of adsorption, $L$
scales as a power law,  $L \propto (\epsilon -\epsilon)^x$. The exponent $x$,
plotted as a function of $N$ in the inset, is negative, therefore, stronger
attraction makes the loops smaller while the mean loop size evidently increases
with growing chain length $N$ which is a finite size effect. The exponent $x$
approaches $-0.96$ in the limit $1/N \rightarrow 0$ - see inset in
Fig~\ref{loop_f0}b. This provides another estimate of the crossover exponent
$\phi$ since $x=1-1/\phi$, according to Eq.(\ref{L_diverge}). Thus we find
$\phi
\approx 0.51 \pm 0.02$. From Fig~\ref{loop_f0}b it is evident that the slope of
the $L$ vs. $(\epsilon -\epsilon)/k_BT$ curves visibly changes as one comes
closer to the CAP. In the immediate vicinity of $\epsilon_c$ the slope is small
and the corresponding estimate for the crossover exponent in this region is
$\phi
\approx 0.63$. One should bear in mind, however, that this is due to the finite
length of the chains used in the simulation which limits the possibility for the
loop size to grow indefinitely, especially at $\epsilon_c$. Therefore, we use
and depict measurements of the slope sufficiently far from the CAP where it
tends to a constant value, indicated by the dashed line in Fig~\ref{loop_f0}b.

\begin{figure}[bht]
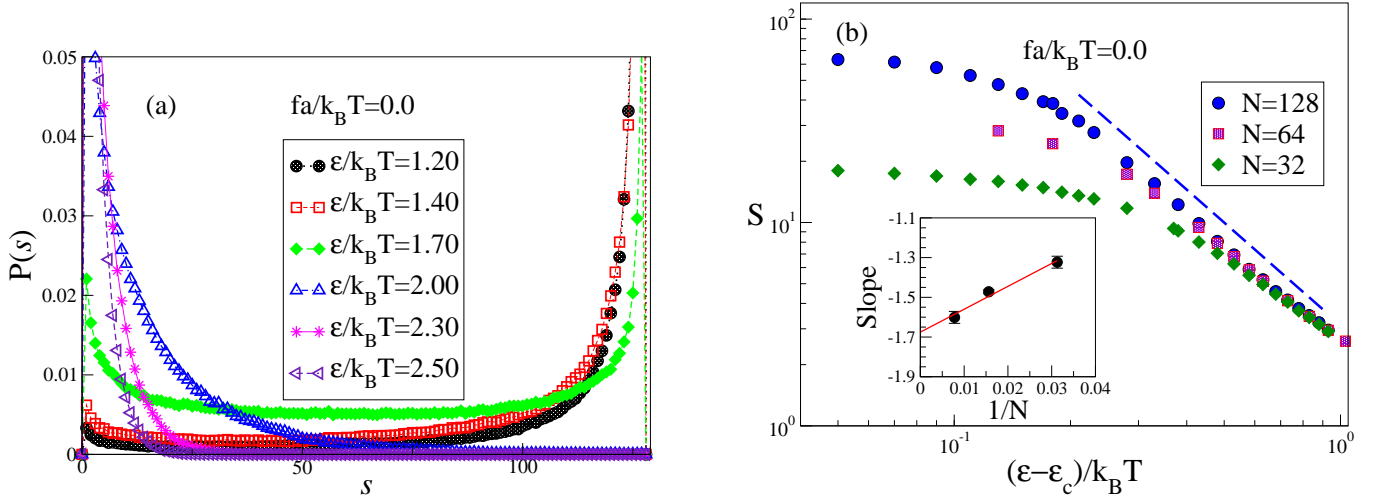

\vspace{0.8cm}
\includegraphics[scale=0.35]{fig9.eps}
\hspace{1cm}
\includegraphics[scale=0.35]{fig10.eps}
\vspace{0.5cm}
\caption{(a)Distribution of the tail size for different surface potentials in a
polymer of length $N = 128$ with no pulling force.  (b) The average tail length
$S$ against $(\epsilon -\epsilon_c)/k_BT$ plotted for various chain lengths in
double logarithmic coordinates. The slopes obtained from these curves are
plotted against $1/N$ in the inset and extrapolated to get the thermodynamic
limit $N\rightarrow \infty$.}
\label{tail_f0}
\vspace{0.5cm}
\end{figure}

In Fig.~\ref{tail_f0}(a) we plot the PDF of the tail size for a chain with
$N=128$ at several strengths of the adsorption potential in the absence of
pulling. An interesting feature of the tail distribution function for $\epsilon
= 1.70$ immediately at the CAP, $\epsilon_c = 1.67$, is the observed {\em
bimodal} character. It means that there are two dominating chain populations,
one with few loops and a long tail, and the other with many loops and a very
short tail. Our simulation result thus confirms the shape of the tail
distribution at criticality, Eq.~(\ref{Tail_distribution__all}), and appears in
excellent agreement with the analytic result, derived earlier by Gorbunov et al.
\cite{Gorbunov}, indicating that in the vicinity of the critical adsorption
point (CAP) chain conformations are either loop- or tail-dominated.

In Fig~\ref{tail_f0}(b) the average tail length, $S$, is plotted against
$(\epsilon -\epsilon_c)/k_BT$. Again, $S$ is found to scale as a power law with
the adhesion strength, $S \propto (\epsilon -\epsilon_c)^y$ where $y$ is
negative, decreases with $N$, and approaches eventually $-1.67$ for $1/N
\rightarrow 0$ . This result can be compared to Eq.(~\ref{Tail_Length}). The
corresponding estimate  of $\phi$  is thus $0.60$.

We turn now to the properties of adsorbed chains in the presence of pulling
force. A remarkable feature of the probability distribution of the order
\begin{figure}[bht]
\vspace{0.8cm}
\includegraphics[scale=0.38]{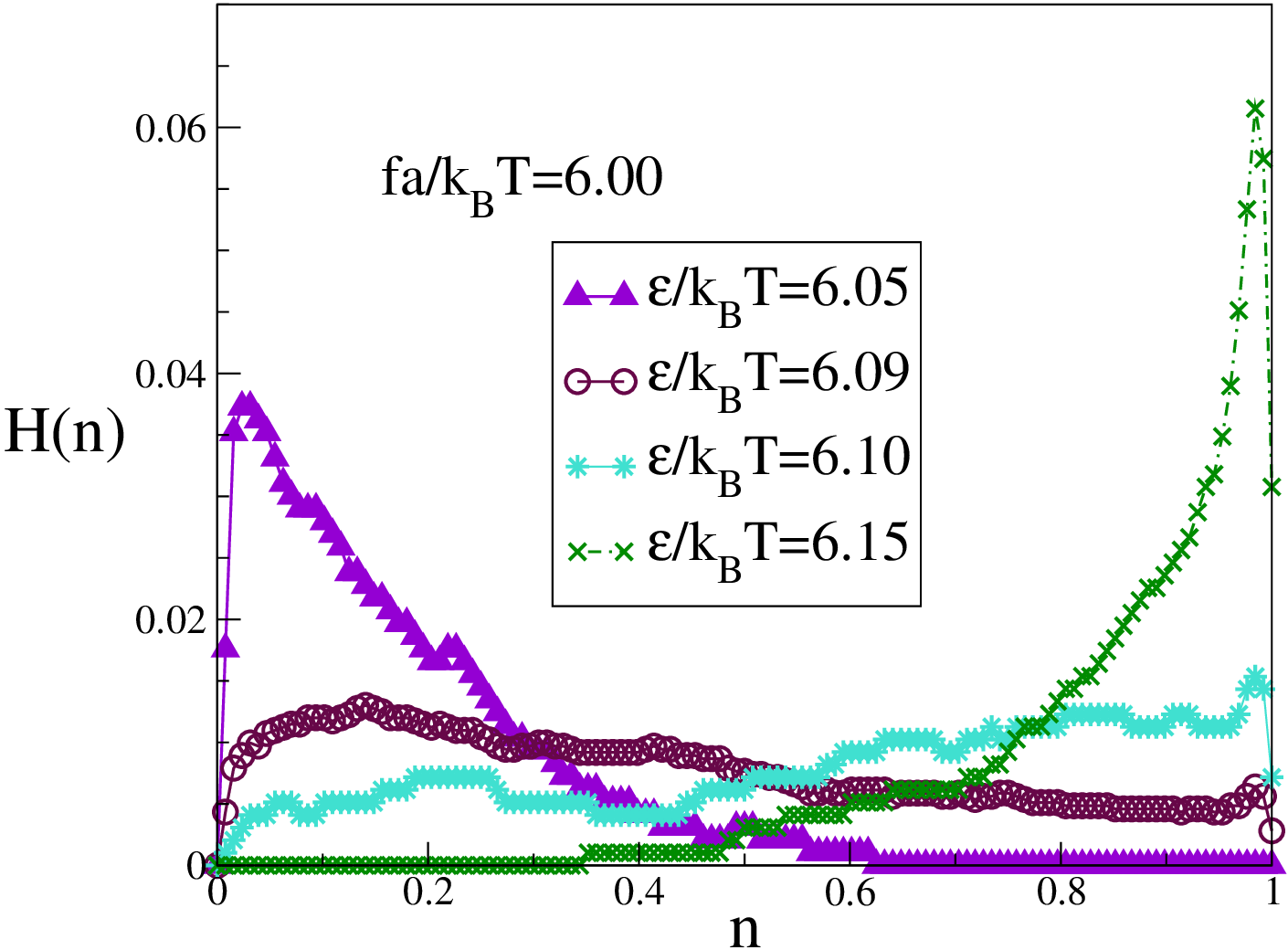}
\caption{Distribution of the order parameter $n$ for a pulling force $fa/k_BT =
6.0$ and different strength of adhesion $\epsilon/k_BT$. The chain length is
$N=128$ and the threshold value of the surface potential for this force is
$\epsilon_D \approx 6.095 \pm 0.03$. The values $\epsilon/k_BT=6.09$ and
$\epsilon/k_BT=6.10$ are on both sides of the detachment line, cf.
Fig.~\ref{Phase-diagram}.}
\label{fig_PDF_n}
\end{figure}
parameter is the absence of a second peak in the vicinity of the critical
strength of adsorption, $\epsilon_D \approx 6.095 \pm 0.03$, which still keeps
the polymer adsorbed at pulling force $fa/k_BT = 6.0$, see Fig.~\ref{fig_PDF_n}.
Somewhat further away from $\epsilon_D$, one observes a clear maximum in the
distribution $H(n)$, indicating a desorbed chain with $n\approx 0.01$ for
$\epsilon = 6.05$, or an almost entirely adsorbed chain with $n\approx 0.99$ for
$\epsilon = 6.15$. This lack of bimodality in the $H(n)$ confirms the dichotomic
nature of the desorption transition which rules out phase coexistence.
 
\begin{figure}[bht]
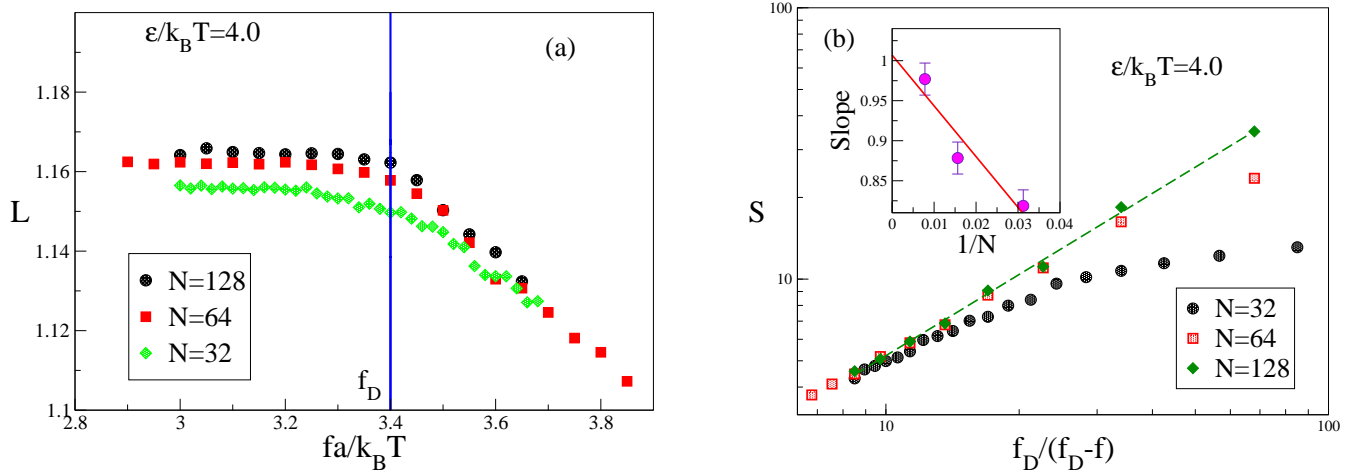

\vspace{0.5cm}
\includegraphics[scale=0.35]{fig16.eps}
\hspace{1cm}
\vspace{0.5cm}
\includegraphics[scale=0.34]{fig15.eps} \caption{(a) The average loop length
plotted against the pulling force $f$ for fixed $\epsilon/k_BT =4.0$. (b) The
average tail length $S$ is plotted against the $(f-f_D)^{-1}$  for various chain
lengths for $\epsilon/k_BT=4.0$ in double logarithmic  coordinates. The inset
shows the extrapolated slope for $N \rightarrow \infty$ go to unity, as
predicted by Eq. (\ref{Tail_Length_Divergency})}.
\label{tail_e2_4}
\end{figure}
In Fig~\ref{tail_e2_4}a, the average loop length, $L$ is plotted against
the external pulling force $f$ for $\epsilon/k_BT =4.0$. For $f$ below the
detachment threshold, $f_D$,  the average loop size appears to be constant
independent of the force. As the force $f$ exceeds $f_D$, the average loop size
decreases in close agreement with the theoretical prediction, shown in
Fig.~\ref{Detachment_theor}a.
\begin{figure}[bht]
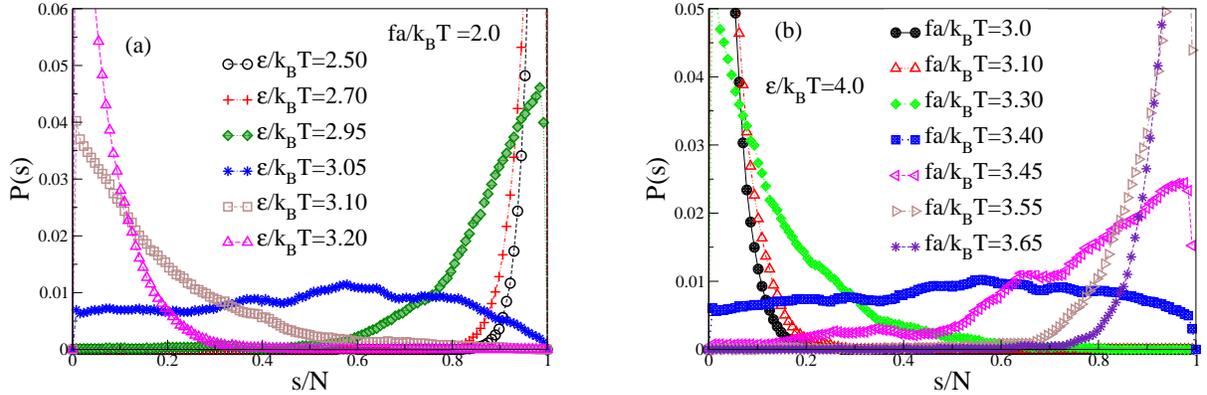

\vspace{0.5cm}
\includegraphics[scale=0.3]{fig20.eps}
\hspace{1cm}
\includegraphics[scale=0.3]{fig17.eps}
\caption{(a) Distribution of the tail size for a pulling force $fa/k_BT = 2.0$
and different strength of adhesion $\epsilon/k_BT$. The chain length is $N=128$.
(b) Distribution of the tail size for different force $f$ at $\epsilon/k_BT =
4.0$.}
\label{tail-e4}
\end{figure}
In Fig~\ref{tail_e2_4}b, the average tail length, $S$, is plotted against the
difference $f_D-f$ for several chain lengths 
at surface adhesion $\epsilon/k_BT=4.0$ in double logarithmic
coordinates. As the applied pulling force $f$ gradually approaches the
threshold force for detachment, $f_D$, the tail gets systematically longer and
comes close to the length of the chain $N$. Evidently, if one takes into
account the finite-size effects which lead to the observed bending of $S
\approx N$ at stronger pulling, the tail $S$ scales as $(f_D-f)^{-w}$. The
exponent $w$ approaches $1.01$ (see inset in Fig~\ref{tail_e2_4}b) at
$\epsilon/k_BT=4.0$. This may be compared to the theoretical prediction of
Eq.~(\ref{Tail_Length_Divergency}) which predicts indeed $w = 1$.

Eventually, in Fig~\ref{tail-e4}(a) the PDF of the tail size $s$ is plotted at
different strengths of the surface potential $\epsilon$ while the force, applied
to the chain end, is held constant, $f = 2.0$.  In contrast, in
Fig~\ref{tail-e4}(b), we display the distribution of tail size $s$ for the case
when the adhesion strength is fixed, $\epsilon/k_BT = 4.0$, whereas the pulling
force $f$ is varied. Both graphs are remarkable in that they reflect the
transition from fully adsorbed polymer, characterized by a sharp peak in the PDF
at vanishing tail sizes, to detached chain when the pulling force exceeds the
threshold $f_D$ and the corresponding PDF is peaked at $s/N \approx 1$. We
emphasize again that although this phase transition of chain detachment is
clearly of first order, no trace of a bimodal distribution in the vicinity of
the transition line can be detected! Thus, the states on both sides of the phase
boundary $f_D(\epsilon)$ {\em cannot coexist simultaneously} which underlines
the peculiar nature of this phase transformation. At this point we should like
to point out, however, that this exotic feature of the  detachment transitions
has meanwhile been established also in the case of the so called {\em escape}
transition of a polymer coil, deformed under the tip of an Atomic Force
Microscope \cite{Sub,Yamakov}. It has been shown rigirously
recently\cite{Klushin}, that despite its first order nature, the escape
transition takes place without phase coexistence. Most probably, this unusual
feature is due to the topological connectivity of polymer chain as
quasi-onedimensional systems.

\section{Summary and Discussion}
\label{sect_summary}

In the present investigation we have studied the force-induced desoption
transition of a polymer chain in contact with an adhesive surface. We treat the
problem within the framework of the Grand Canonical Ensemble approach and
derive analytic expressions for the various conformational building blocks,
characterizing the structure of an adsorbed linear polymer chain, subject to
pulling force of fixed strength. Closed analytic expressions for the
fraction of adsorbed segments (i.e., the order parameter of the
desorption transition) and for probability distributions of trains, loops and
tails have been derived along with expressions for the corresponding first
moments in terms of the surface potential intensity both with and without
external force. As expected, all these conformational properties and their
variation with the proximity to the CAP are governed by a crossover exponent
$\phi$.

A central result in the present work is the calculation of $\phi$ using the
approach of Kafri et al. \cite{Kafri} which provides insight into the background
of the existing controversial reports about its numeric value. We demonstrate
that the value of $\phi$ may vary within the interval $0.39 \le \phi \le 0.6$,
depending on the possibility of a single loop to interact with the neighboring
loops in the adsorbed polymer. Since this range is model-dependent, one should
not be surprised that different models produce different estimates of $\phi$ in
this interval.

A comparison with the results from extensive Monte Carlo simulations
demonstrates the good agreement between theoretic predictions and simulation
data.

In particular, we verify the gradual transition of the PDF of loops from
power-law to exponential decay as one moves away from the critical adsorption
point to stronger adsorption. We demonstrate that for vanishing pulling force,
$f \rightarrow 0$, the mean loop size, $L \propto (\epsilon -
\epsilon_c)^{1-\frac{1}{\phi}}$, and the mean tail size, $S \propto (\epsilon -
\epsilon_c)^{\frac{1}{\phi}}$, diverge when one comes close to the CAP. In
contrast, for a non-zero pulling force, $f \ne 0$, we show that the loops on the
average get smaller with growing force while close to the detachment threshold,
$f \approx f_D$, the tail length diverges as $S \propto (1-\frac{f}{f_D})^{-1}$.

Eventually, we derive the overall phase diagram of the force-induced desorption
transition for a linear self-avoiding polymer chain and demonstrate its {\em
reentrant} character when plotted in terms of detachment force $f_D$ against
system temperature $T$. We find that despite being of first order, the
force-induced phase transition of polymer desorption is dichotomic in its
nature, that is, no phase coexistence and no metastable states exist. This
unusual feature of the phase transformation is unambiguously supported by our
simulation data, e.g., through the comparison of the order parameter probability
distributions on both sides in the immediate vicinity of the detachment line
whereby no double-peaked structure is detected. 

Finally, we should like to to emphasize that while the present investigation
will hopefully shed new light on the force-induced desorption transition of a
linear polymer from a sticky surface, a lot more work is needed before a
comprehensive understanding of this phenomenon is achieved. In this work
simulations have been carried out within the framework of a {\em constant
force} ensemble. In their comprehensive treatment of the problem,
however, Skvortsov et al. \cite{SKB} have shown that one may well work in the
{\em constant height} ensemble whereby one uses the end-monomer $h$-position as
an independent parameter and measures the force, exerted by the chain on the end
monomer. Notwithstanding the equivalence of both ensembles, some quantities
behave differently in each ensemble and this becomes evident only if one
presents the salient features of the system behavior in each particular
ensemble. Thus in the fixed-height ensemble, a different and rather interesting
thermodynamic behavior of the measured mean detachment force and of the fraction
of adsorbed segments agains $h$ is expected to be observed. Typically one then
observes a constant force plateau while the height of the chain end monomer is
varied. Such a behavior can be inferred even within the fixed-force ensemble.
Indeed, as shown in Fig.~\ref{tail-e4}b  the PDF $P(s)$ is practically
flat for the critical detachment force $f_D = 3.40$, meaning that all chain end
heights are equally probable at this particular force. The latter is equivalent
to a constant-force plateau in the fixed-height ensemble. A verification by
computer experiment is among our tasks in the immediate future as well as a
study of the so far unexplored  kinetics of chain detachment.
 
\section{Acknowledgments}
We are indebted to A. Skvortsov, L. Klushin, J.-U. Sommer, and K. Binder for
useful discussions during the preparation of this work. A.~Milchev thanks the
Max-Planck Institute for Polymer Research in Mainz, Germany, for hospitality
during his visit in the institute. A.~Milchev and V.~Rostiashvili acknowledge
support from the Deutsche Forschungsgemeinschaft (DFG), grant No. SFB 625/B4.
\begin{appendix}
\section{Properties of the polylog function}
\label{sect_polylog}
The polylog function $\Phi (\alpha, z)$ is defined by the series
\begin{eqnarray}
 \Phi (\alpha, z) = \sum_{k=1}^{\infty} \: \frac{1}{k^{\alpha}} \; z^k
\label{Polylog}
\end{eqnarray}
which converges at $|z| < 1$. From the definition, Eq. (\ref{Polylog}), one
immediately obtains
\begin{eqnarray}
 z \: \frac{d}{d z} \: \Phi(\alpha, z) = \Phi(\alpha - 1, z)
\label{Property}
\end{eqnarray}

The calculation of the series Eq. (\ref{Polylog}) (see Sec. 1.11 in ref.
\cite{Erdelyi}) gives
\begin{eqnarray}
 \Phi(\alpha, z) = \Gamma(1-\alpha) \left[ \ln \left( \frac{1}{z}\right)
\right]^{\alpha-1} + \sum_{r=0}^{\infty} \: \zeta (\alpha - r) \frac{(\ln
z)^{r}}{r !}
\label{Phy}
\end{eqnarray}
where $\Gamma(x)$ is the gamma-function, $\zeta (x)$ is the Riemann
zeta-function, and the exponent $\alpha$ is noninteger, i.e. $\alpha \neq 1, 2,
3, \dots$

Consider now the case of integer values of $\alpha$. The gumma-function
$\Gamma(x)$ has poles at all negative integer arguments whereas the pole of
$\zeta (x)$ is placed at $x=1$. One may write $\alpha = m+\delta$ where $m$ is a
positive integer and $\delta \rightarrow 0$. Then in the vicinity of the poles
the gamma- and zeta-functions can be rewritten as
\begin{eqnarray}
\Gamma(1 - m - \delta) &=& \frac{(-1)^m}{(m-1)!} \: \left[ \frac{1}{\delta} -
\psi(m) + {\cal O}(\delta)\right] \nonumber\\
 \zeta (1 + \delta) &=& \left[ \frac{1}{\delta} - \psi(1) + {\cal
O}(\delta)\right]
\label{Gamma_Zeta}
\end{eqnarray}
where $\psi(x)$ is the digamma function (or $\psi$-function) defined as the
logarithmic derivative of the gamma-function, $\psi(x) = d \ln \Gamma(x)/d x$.
One should also take into account that
\begin{eqnarray}
\left[ \ln \left( \frac{1}{z}\right) \right]^{\delta} = 1 + \delta \ln \left[
\ln \left( \frac{1}{z}\right) \right] + {\cal O}(\delta)
\label{Log_Log}
\end{eqnarray}
After taking into account Eqs.(\ref{Gamma_Zeta}) and (\ref{Log_Log}) in Eq.
(\ref{Phy}) and due to the cancellation of poles in the gamma- and
zeta-functions at small values of $\delta$ the polylog function,
Eq.(\ref{Polylog}) becomes
\cite{Erdelyi}
\begin{eqnarray}
\Phi(m, z) = \frac{(\ln z)^{m-1}}{(m-1)!} \: \left[ \psi(m) - \psi(1) - \ln \ln
\left( \frac{1}{z}\right) \right]  + \mathop{{\sum}'}\limits_{r=0}^{\infty} \:
\zeta (m - r) \;  \frac{(\ln z))^{r}}{r!}
\label{Integer}
\end{eqnarray}
where the prime indicates that the term $r = m - 1$ is to be omitted.

We are interested in the behavior of $\Phi(\alpha, z)$ at $z \rightarrow 1$. In
this case $\ln (1/z) = - \ln [1- (1-z)] \approx (1-z)$. At $\alpha < 1$, the
main contribution comes from the first term in Eq. (\ref{Phy}), i.e.
\begin{eqnarray}
 \Phi(\alpha, z) \approx \frac{\Gamma(1 - \alpha)}{(1 - z)^{1-\alpha}}
\label{Alpha_Less}
\end{eqnarray}

At $\alpha = 1$  and $z \rightarrow 1$, and making use of Eq. (\ref{Integer}),
one obtains
\begin{eqnarray}
 \Phi(1, z) \approx -\ln \ln \left(\frac{1}{z}\right) \approx \ln
\left(\frac{1}{1
- z}\right)
\label{Alpha_One}
\end{eqnarray}
Finally, at $\alpha > 1$ the polylog function $\Phi(\alpha, z)$ has no
singularity at $z \rightarrow 1$ and Eq.(\ref{Phy}) results in the following
expansion
\begin{eqnarray}
 \Phi(\alpha, z) \approx \zeta (\alpha) + \Gamma(1-\alpha) (1 - z)^{\alpha-1} -
\zeta(\alpha-1) (1 - z)  + \dots
\label{Alpha_More}
\end{eqnarray}
In a bit more specific case when $1 < \alpha < 2$ we will use the well known
relationship $\Gamma(1 - \alpha) = -\pi/[\Gamma(\alpha) |\sin ( \pi \alpha)|]$ 
so that
\begin{eqnarray}
 \Phi(\alpha, z) \approx \zeta (\alpha) -  \frac{\pi}{\Gamma(\alpha) |\sin(\pi
\alpha)|} (1 - z)^{\alpha-1} - \zeta(\alpha-1) (1 - z)  + \dots
\label{Alpha_More_One}
\end{eqnarray}

Taking into account the Eqs.(\ref{Alpha_Less}), (\ref{Alpha_One}) and
(\ref{Alpha_More_One}), the expression for the polylog function at $z
\rightarrow 1$ reads
\begin{eqnarray}
\Phi(\alpha, z) \approx \begin{cases}
                        \frac{\Gamma(1 - \alpha)}{(1 - z)^{1-\alpha}}, \quad 
&\mbox{\rm at} \quad  \alpha < 1\\
                          \\
\ln \left(\frac{1}{1 - z}\right),  \quad  &\mbox{\rm at} \quad  \alpha = 1\\
\\
\zeta (\alpha) - a_{\alpha} (1 - z)^{\alpha-1} - b_{\alpha} (1 - z)+ \dots ,
\quad  &\mbox{\rm at} \quad  1 < \alpha < 2
                       \end{cases}
\label{Cases}
\end{eqnarray}
where the coefficients $a_{\alpha} = \pi/{\Gamma(\alpha) |\sin (\pi \alpha)|}$
and $b_{\alpha} = \zeta(\alpha - 1)$.

\end{appendix}

\begin{figure}
\includegraphics[scale=0.6]{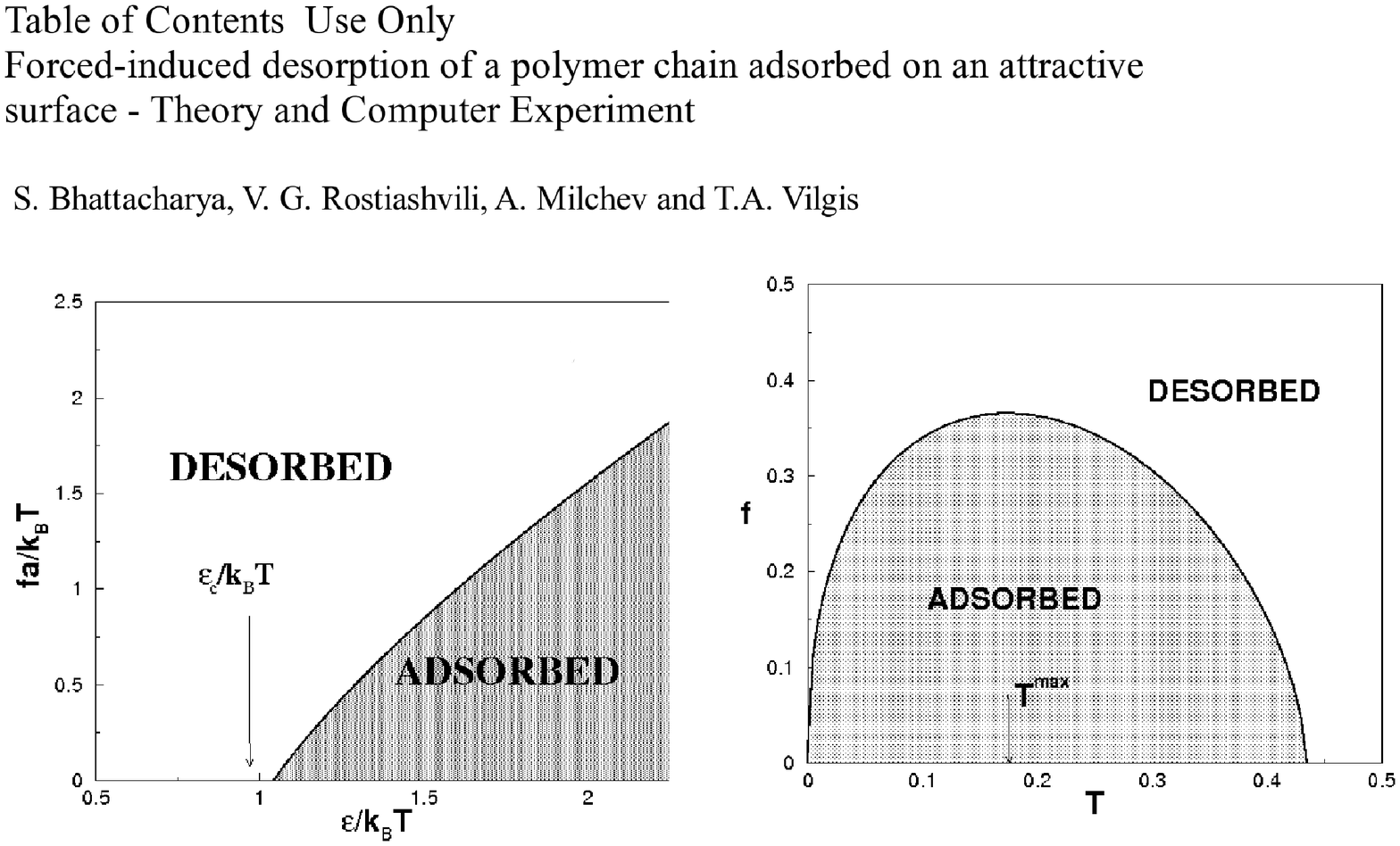}
\end{figure}

\end{document}